\documentclass[conference]{IEEEtran}
\ifCLASSINFOpdf
\else
\fi
\usepackage{graphicx}
\usepackage{url}
\usepackage{float}
\usepackage{amssymb,amsmath}
\usepackage{stfloats}
\usepackage{color}
\usepackage{subfig}
\usepackage{amsthm}
\usepackage{comment}
\usepackage{paralist}
\usepackage{epstopdf}
\usepackage[nofillcomment,ruled,linesnumbered]{algorithm2e}
\SetAlFnt{\footnotesize}
\SetAlCapSkip{0.5ex}
\usepackage{tikz}
\newcommand{\ra}{\rightarrow}
\newcommand{\la}{\leftarrow}
\newcommand{\pf}{IEEEproof}
\newcommand{\remove}[1]{}

\newcommand{\join}{\, \sqcup \,}
\newcommand{\meet}{\, \sqcap \,}

\newcommand{\cuts}[1]{\mathcal{C}(#1)}
\newtheorem{lem}{Lemma}[]

\newtheorem{thm}{Theorem}[]
\newtheorem{defn}{Definition}[]
\ifCLASSINFOpdf
\graphicspath{{./figs/}}
\DeclareGraphicsExtensions{.eps}
\else
\graphicspath{{./figs/}}
\DeclareGraphicsExtensions{.eps}
\fi

\begin{document}
\title{A Distributed Abstraction Algorithm for Online Predicate Detection}

\author{\IEEEauthorblockN{Himanshu Chauhan \quad Vijay K. Garg}
\IEEEauthorblockA{Dept. of Electrical and Computer Engineering\\
The University of Texas at Austin\\
himanshu@utexas.edu~~garg@ece.utexas.edu}
\and
\IEEEauthorblockN{Aravind Natarajan \quad Neeraj Mittal}
\IEEEauthorblockA{Dept. of Computer Science\\
The University of Texas at Dallas\\
aravindn@utdallas.edu~~neerajm@utdallas.edu}
}


%Telephone: (800) 555--1212, Fax: (888) 555--1212}


\maketitle

\begin{abstract}
%\boldmath
Analyzing a distributed computation is 
a hard problem in general due to the combinatorial 
explosion in the size of the state-space with the 
number of processes in the system. By abstracting the computation, 
unnecessary state explorations can be avoided. Computation slicing is an approach for 
abstracting distributed computations with respect to a given predicate. 
We focus on regular predicates, a family of predicates that covers many commonly used predicates for runtime verification. 
The existing algorithms for computation slicing are 
centralized -- a single process is 
responsible for computing the slice in either offline
or online manner. In this paper, we present first
distributed online algorithm for computing the slice of 
a distributed computation with respect to a regular 
predicate. Our algorithm distributes the work and storage requirements
across the system, thus reducing the space 
and computation complexity per process. 
%In addition, for conjunctive predicates, our algorithm also reduces the message load per process.  
\end{abstract}


% \fi
\IEEEpeerreviewmaketitle

\section{Introduction}

\remove{
Verifying system states and global
properties is a critical aspect of maintaining consistency and detecting constraint 
violations for reliable distributed system. }
Global predicate detection \cite{CooMar:1991:WPDD} for runtime verification
is an important technique for detecting violations of 
invariants for debugging and fault-tolerance in 
distributed systems. 
It is a challenging task on a large system
with a large number of processes due to the combinatorial explosion of the 
state space. The predicate detection problem is not only applicable to conventional
distributed systems, but 
also to multicore computing.  
 With growing popularity of large number of CPU-cores on 
processing chips \cite{tilera}, some manufacturers are exploring the 
distributed computing model on chip with no shared 
memory between the cores, and information 
exchange between the cores only using message passing \cite{rotta2012memory}.
Recent research efforts 
\cite{PetrovicSRS12}  have shown that with sufficiently fast on-chip networking support, 
such a message passing based model can provide significantly fast performance for some specific computational tasks. 
With emergence of these trends, techniques in predicate detection for distributed systems can also 
 be useful for systems with large number of cores.

Multiple algorithms 
have been proposed in literature for detection of global predicates in both offline and online manner (e.g. \cite{CooMar:1991:WPDD,MitSen+:2007:TPDS,AlaVen:2001:TSE}).
 Online predicate detection is important for many system models: 
continuous services (such as web-servers), collection of continuous 
observations (such as sensor-networks), and parallel search operations 
on large clusters. However,  
performing predicate detection in a manner that is oblivious to the 
structure of the predicate can lead to large runtime, and high 
memory overhead. 
The approach of using mathematical abstractions for designing and analyzing computational tasks has proved to be significantly advantageous in modern computing. 
In the context of predicate detection, and runtime verification, the 
problem of abstraction can be viewed as the problem of taking a distributed computation as input and
outputting a smaller distributed computation that abstracts out parts that are not relevant 
to the predicate under consideration. The abstract computation may be exponentially smaller than the
original computation resulting in significant savings in predicate detection time.

\emph{Computation slicing} is an abstraction technique for efficiently finding all global 
states, of a distributed computation, that satisfy a given global 
predicate, without explicitly enumerating all such global states 
\cite{MitSen+:2007:TPDS}.
The {\em slice} of a computation with respect 
to a predicate is a sub-computation that satisfies the following properties:
(a)  it contains all global states of the computation for which the 
predicate evaluates to true, and (b) of all the sub-computations that satisfy condition (a), it has the 
least number of global states. 
As an illustration, consider the computation shown in
Fig.~\ref{fig:firstcomputation}\subref{fig:firstex}. The computation consists of three
processes $P_1$, $P_2$, and $P_3$ hosting integer variables $x_1$,
$x_2$, and $x_3$, respectively. An event, represented by a 
circle is labeled with the value of the 
variable immediately after the event is executed. 

Suppose we want to determine whether the
property (or the predicate) $(x_1 * x_2 + x_3 < 5)$ $\wedge$ $(x_1
\geq 1)$ $\wedge$ $(x_3 \leq 3)$ ever holds in the computation. In
other words, does there exist a global state of the computation that
satisfies the predicate? The predicate could represent the violation
of an invariant. Without computation slicing, we are forced to examine
all global states of the computation, twenty-eight in total, to
ascertain whether some global state satisfies the
predicate. Alternatively, we can compute a slice of the computation automatically
with respect to the predicate $(x_1 \geq 1)
\wedge (x_3 \leq 3)$ as shown in Fig.~\ref{fig:firstcomputation}\subref{fig:firstexslice}. 

We can now restrict our search to the
global states of the slice, which are only six in number,
namely:\\
$\{
a, e, f, u, v \}, \{ a, e, f, u, v, b \}, \{ a, e, f, u, v, w \},$\\
$\{ a, e, f, u, v, b, w \}, 
\{ a, e, f, u, v, w, g \}$, and $\{ a, e,
f, u, v, b, w, g
\}$. 

The slice has much fewer global states than the computation
itself---exponentially smaller in many cases---resulting in
substantial savings.
\tikzstyle{place}=[circle,draw=black,fill=black,thick,inner sep=0pt,minimum size=2mm]
\tikzstyle{nsplace}=[circle,draw=black,thick,inner sep=0pt,minimum size=2mm]
\tikzstyle{place1}=[circle,draw=black,thick,inner sep=0pt,minimum size=3mm]
\tikzstyle{inner}=[circle,draw=black,fill=black,thick,inner sep=0pt,minimum size=1.25mm]
\tikzstyle{satlatnode}=[ellipse,draw=black,fill=black!20,thick,minimum size=5mm]
\tikzstyle{nsatlatnode}=[ellipse,style=dashed,thick]
\tikzstyle{satblock} = [rectangle, draw=gray, thin, fill=black!20,
text width=2.5em, text centered, rounded corners, minimum height=2em]
\tikzstyle{nsatblock} = [rectangle, draw=gray, thin,
text width=2.5em, text centered, rounded corners, minimum height=2em]
\tikzstyle{jbblock} = [rectangle, draw=black, thick, fill=black!20,
text width=2.5em, text centered, minimum height=2em]

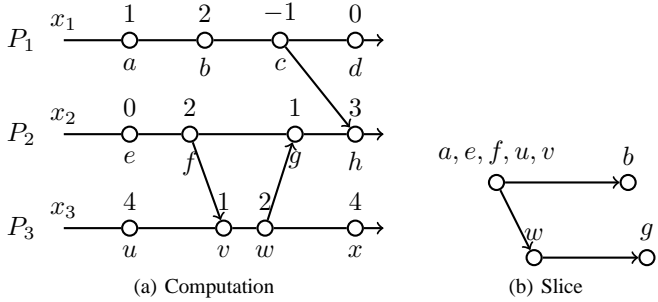
\begin{figure}
\centering
\subfloat[Computation]{
\begin{tikzpicture}
\node at ( 0,1.25) [nsplace] [label=below:$a$,label=above:$1$] (a) {};
\node at ( 1,1.25) [nsplace] [label=below:$b$,label=above:$2$] (b) {};
\node at ( 2,1.25) [nsplace] [label=below:$c$,label=above:$-1$] (c) {};
\node at ( 3,1.25) [nsplace] [label=below:$d$,label=above:$0$] (d) {};
%\node at ( 0,1) [place] {};
\node at ( 0,0) [nsplace] [label=below:$e$,label=above:$0$] (e) {};
\node at ( 0.8,0) [nsplace] [label=below:$f$,label=above:$2$](f) {};
\node at ( 2.2,0) [nsplace] [label=below:$g$,label=above:$1$] (g) {};
\node at ( 3,0) [nsplace] [label=below:$h$,label=above:$3$] (h) {};
\node at ( 0,-1.25) [nsplace] [label=below:$u$,label=above:$4$] (u) {};
\node at ( 1.25,-1.25) [nsplace] [label=below:$v$,label=above:$1$](v) {};
\node at ( 1.8,-1.25) [nsplace] [label=below:$w$,label=above:$2$] (w) {};
\node at ( 3,-1.25) [nsplace] [label=below:$x$,label=above:$4$] (x) {};

%
%\node at ( 1,1) [rectangle,draw] {};
%\node at (-1,1) [rectangle,draw] {};
%edge [pre] (c)
%edge [post] (f)
\node at (3.5, 1.25) [] (end1) {};
\node at (3.5, 0) [] (end2) {};
\node at (3.5, -1.25) [] (end3) {};

\draw [thick] (b.west) -- (a.east);
\draw [thick] (c.west) -- (b.east);
\draw [thick] (c.east) -- (d.west);
\draw [thick] (g.east) -- (h.west);
\draw [thick,->] (d.east) -- (end1.west);
\draw [thick,->] (h.east) -- (end2.west);

\draw [thick] (f.west) -- (e.east);
\draw [thick] (g.west) -- (f.east);

\draw [thick] (v.west) -- (u.east);
\draw [thick] (w.west) -- (v.east);
\draw [thick] (x.west) -- (w.east);
\draw [thick,->] (x.east) -- (end3.west);

\node at (-1, 1.25) [label=left:$P_{1}$] (bot1) {};
\node at (-1, 0) [label=left:$P_{2}$] (bot2) {};
\node at (-1, -1.25) [label=left:$P_{3}$] (bot3) {};

\draw [thick] (a.west) -- (bot1.east)
node [above]
{
$x_{1}$
};
\draw [thick] (e.west) -- (bot2.east)
node [above]
{
$x_{2}$
};
\draw [thick] (u.west) -- (bot3.east)
node [above]
{
$x_{3}$
};
\draw [thick,->] (c) -- (h);
\draw [thick,->] (f) -- (v);
\draw [thick,->] (w) -- (g);
\end{tikzpicture}
%\caption{Computation }
\label{fig:firstex}
}
\subfloat[Slice]{
\begin{tikzpicture}
%\remove{
\node at ( 0,1) [nsplace] [label={above:$a,e,f,u,v$}] (a) {};
\node at ( 1.75,1) [nsplace] [label=above:$b$] (b) {};
\node at ( 0.5,0) [nsplace] [label=above:$w$] (w) {};
\node at ( 2,0) [nsplace] [label=above:$g$] (g) {};
\draw [thick,->] (a) -- (b);
\draw [thick,->] (a) -- (w);
\draw [thick,->] (w) -- (g);
%}
\end{tikzpicture}
\label{fig:firstexslice}
}
\caption{A Computation, and its slice with respect to predicate $(x_1 \geq 1) \wedge (x_2 \leq 3)$}
\label{fig:firstcomputation}
\end{figure}

\remove{
\begin{figure}[htbp]
\includegraphics[width=5.5in]{motivation1.eps}
%\centerline{\epsfbox{motivation1.eps}}
%\caption{\label{fig:motivation1} (a) A computation and (b) its slice with
%respect to $(x_1 \geq 1) \wedge (x_3 \leq 3)$.}
\end{figure}
}

In this paper, 
we focus on abstracting distributed computations with respect to {\em regular} predicates (defined in Sec. \ref{sec:model}). The family of {\em regular} predicates 
contains many useful predicates that are often used for runtime verification 
in distributed systems. Some such predicates are:\\
{\bf Conjunctive Predicates}: Global predicates which are conjunctions of local predicates.
For example,
predicates of the form, $B = (l_1 \geq x_1 \geq u_1) \wedge (l_2 \geq x_2 \geq u_2) \wedge
\ldots \wedge (l_n \geq x_n \geq u_n)$, where $x_i$ is the local variable on 
process $P_i$, and $l_i$, $u_i$ are constants, are conjunctive predicates. Some useful verification predicates
that are in conjunctive form are: detecting mutual exclusion violation in 
pairwise manner, pairwise data-race detection, detecting if each process 
has executed some instruction, etc.\\ 
\hspace{0.5pt}
{\bf Monotonic Channel Predicates} \cite{Gar:2002:Book}: 
Some examples are: all messages have been delivered 
(or all channels are empty), at least $k$ messages have been sent/received, 
there are at most $k$ messages in transit between two processes, the leader has sent all ``prepare to commit'' messages, etc.  

\remove{By using {\em slicing} the problem of predicate detection can be solved 
in efficient manner for regular (and linear) predicates.} 
Centralized
offline \cite{MitGar:2005:DC} and 
online \cite{SenGar:2007:TC} algorithms for {\em slicing}
based predicate detection have been presented previously. In this paper, we present the first 
{\em distributed online} slicing algorithm for
{\em regular} predicates in 
distributed systems.

\subsection{Challenges and Contributions}
\label{sec:contri}
Computing the slice of a computation with respect to a general predicate is 
a NP-Hard problem in general \cite{MitGar:2005:DC}. Many classes of predicates have been 
identified for which the slice can be computed efficiently in polynomial 
time (\emph{e.g.}, regular predicates, co-regular predicates, linear predicates, relational predicates, stable predicates) 
\cite{MitGar:2005:DC,MitSen+:2007:TPDS,SenGar:2007:TC,OgaGar:2007:DISC}. 
However, the existing slicing algorithms are \emph{centralized} in nature. 
The slice is computed by a single \emph{slicer} process
that examines every relevant event in the computation. The centralized algorithms may 
be \emph{offline}, where all events are known a priori, or \emph{online}, 
where the slice is updated incrementally with the arrival of every new relevant event. For systems with large number
of processes, such centralized algorithms require a single process to perform high number 
of computations, and to store very large data. In comparison, a distributed online algorithm significantly
reduces the per process costs for both computation and storage. 
Distributed algorithms are generally faster 
in comparison to centralized algorithms and allow division of tasks among multiple 
processes. Additionally, for predicate detection, the centralized online algorithm requires at least one 
message to the slicer process for every relevant event in the 
computation, resulting in a bottleneck at the slicer process. 

A method of devising a distributed algorithm from a centralized 
algorithm is to 
decompose the centralized execution steps into multiple steps to be
executed by each process
independently. However, for performing online abstraction using computation slicing,
 such an incremental modification would lead to a large number of messages,
 computational steps, and memory overhead. The reason for this inefficiency is that by 
 directly decomposing the steps of the centralized online algorithm, the 
 {\em slicing} computation would require each process to send 
 its local state information to all the other processes whenever the 
 local state (or state interval) is updated. In addition, 
 only splitting the centralized algorithm across all processes 
 leads to a distributed algorithm that wastes significant
 computational time as multiple processes may end 
 up visiting (and enumerating) the same global state. Thus, the task of devising an efficient 
distributed algorithm for {\em slicing} 
is non-trivial. In this paper, we propose a distributed algorithm that exploits not 
only the nature of
the predicates, but also the collective knowledge across 
processes. 
The optimized version of our algorithm reduces the required storage 
per {\em slicing} process, and computational workload per
{\em slicing} process by $O(n)$. An experimental evaluation 
of our proposed approach with the centralized approach 
can be found in the extended version of this paper \cite{arxfull}.

\remove{
We 
discuss these issues and their solutions in detail in Section~\ref{sec:distAlg}. 
}

\subsection{Applications}
\label{conc}
Our algorithm is useful for global predicate detection. Suppose the predicate
 $B$ is of the form $B_1 \wedge B_2$, where $B_1$ is regular but $B_2$ is not. 
We can use our algorithm to slice with respect to $B_1$ to 
reduce the time and space complexity of the global predicate detection. Instead of searching for 
the global state that satisfies $B$ in the original computation,
with the distributed algorithm we can search the global states in the slice for $B_1$.
For example, the Cooper-Marzullo algorithm traverses the lattice of global states in an online manner
\cite{CooMar:1991:WPDD}, which can be quite expensive.  By running our algorithm together with Cooper-Marzullo algorithm, the space and time
complexity of predicate detection is reduced significantly (possibly exponentially) for predicates
in the above mentioned form. 

Our algorithm is also useful for recovery of distributed programs based on checkpointing.
For fault-tolerance, we may want to restore a distributed computation to a checkpoint
which satisfies the required properties such as ``all channels are empty", and ``all processes are in
some states that have been saved on storage". If we compute the slice of the computation in 
an online fashion, then
on a fault, processes can restore the global state that corresponds to the maximum of the last vector
of the slice at each surviving process. This global state is consistent as well as recoverable
from the storage.

\section{Background: Regular Predicates and Slicing}
\label{sec:model}
\subsection{Model}
We assume a loosely coupled asynchronous message passing system, consisting 
of $n$ reliable processes (that do not fail), denoted by $\{P_{1}, P_{2}, 
\ldots, P_{n} \}$, without any shared memory or global clock. Channels are assumed to be FIFO, and loss-less.  
In our model, each local state change is considered an 
event; and every message activity (send or receive) is 
also represented by a new event. We assume that the 
computation being analyzed does not deadlock. 

A distributed computation  is
modeled as a partial order on a set of events \cite{Lam:1978:CACM}, given by Lamport's \emph{happened-before} ($\ra$) relation 
\cite{Lam:1978:CACM}.
We use $(E,\ra)$ to denote the distributed computation on a set of events $E$.
Mattern \cite{Mat:1989:WDAG} and Fidge \cite{Fid:1988:ACSC} proposed {\em vector clocks}, an approach for time-stamping events in a computation such that 
the happened-before relation can be tracked. If $V$ denotes the vector clock for an event $e$ in a distributed computation, then for any event $f$ in
the computation: $
e \ra f \Leftrightarrow e.V < f.V$.
For any pair of events $e$ and $f$ such that $e \not \ra 
f \wedge f \not \ra e$, $e$ and $f$ are said to be
concurrent, and this relation is denoted by 
$e || f$.  
Fig.~\ref{fig:computation}\subref{fig:comp} shows a 
sample distributed computation, and its corresponding 
vector clock representation is presented in Fig.~\ref{fig:computation}\subref{fig:vecclocks}. 
\tikzstyle{place}=[circle,draw=black,fill=black,thick,inner sep=0pt,minimum size=2mm]
\tikzstyle{nsplace}=[circle,draw=black,thick,inner sep=0pt,minimum size=2mm]
\tikzstyle{place1}=[circle,draw=black,thick,inner sep=0pt,minimum size=3mm]
\tikzstyle{inner}=[circle,draw=black,fill=black,thick,inner sep=0pt,minimum size=1.25mm]
\tikzstyle{satlatnode}=[ellipse,draw=black,fill=black!20,thick,minimum size=5mm]
\tikzstyle{nsatlatnode}=[ellipse,style=dashed,thick]
\tikzstyle{satblock} = [rectangle, draw=gray, thin, fill=black!20,
text width=2.5em, text centered, rounded corners, minimum height=2em]
\tikzstyle{nsatblock} = [rectangle, draw=gray, thin,
text width=2.5em, text centered, rounded corners, minimum height=2em]
\tikzstyle{jbblock} = [rectangle, draw=black, thick, fill=black!20,
text width=2.5em, text centered, minimum height=2em]
\begin{figure*}[!bth]
\subfloat[Computation]{
\begin{tikzpicture}
\node at ( 0,1) [nsplace] [label=below:$a$] (a) {};
\node at ( 1,1) [nsplace] [label=below:$b$] (b) {};
\node at ( 2.5,1) [nsplace] [label=below:$c$] (c) {};
%\node at ( 0,1) [place] {};
\node at ( 0,0) [nsplace] [label=below:$e$] (e) {};
\node at ( 1.5,0) [nsplace] [label=below:$f$](f) {};
\node at ( 3,0) [nsplace] [label=below:$g$] (g) {};
%
%
%\node at ( 1,1) [rectangle,draw] {};
%\node at (-1,1) [rectangle,draw] {};
%edge [pre] (c)
%edge [post] (f)
\node at (3.5, 1) [] (end1) {};
\node at (3.5, 0) [] (end2) {};

\draw [thick] (b.west) -- (a.east);
\draw [thick] (c.west) -- (b.east);
\draw [thick,->] (c.east) -- (end1.west);
\draw [thick,->] (g.east) -- (end2.west);

\draw [thick] (f.west) -- (e.east);
\draw [thick] (g.west) -- (f.east);

\node at (-1, 1) [label=left:$P_{1}$] (bot1) {};
\node at (-1, 0) [label=left:$P_{2}$] (bot2) {};

\draw [thick] (a.west) -- (bot1.east)
\remove{
node [above]
{
$x_{1}$
}};
\draw [thick] (e.west) -- (bot2.east)
\remove{
node [above]
{
$x_{2}$
}};
\draw [thick,->] (b) -- (f);
\end{tikzpicture}
%\caption{Computation }
\label{fig:comp}
}
%----
\subfloat[Vector Clock Values of Events]{
\begin{tikzpicture}
\node at ( 0,1) [nsplace] [label=below:$a$,label={above:$[1,0]$}] (a) {};
\node at ( 1,1) [nsplace] [label=below:$b$,label={above:$[2,0]$}] (b) {};
\node at ( 2.5,1) [nsplace] [label=below:$c$,label={above:$[3,0]$}] (c) {};
%\node at ( 0,1) [place] {};
\node at ( 0,0) [nsplace] [label=below:$e$,label={above:$[0,1]$}] (e) {};
\node at ( 1.5,0) [nsplace] [label=below:$f$,label={above:$[2,2]$}] (f) {};
\node at ( 3,0) [nsplace] [label=below:$g$,label={above:$[2,3]$}] (g) {};
%
%
%\node at ( 1,1) [rectangle,draw] {};
%\node at (-1,1) [rectangle,draw] {};
%edge [pre] (c)
%edge [post] (f)
\node at (3.5, 1) [] (end1) {};
\node at (3.5, 0) [] (end2) {};

\draw [thick] (b.west) -- (a.east);
\draw [thick] (c.west) -- (b.east);
\draw [thick,->] (c.east) -- (end1.west);
\draw [thick,->] (g.east) -- (end2.west);

\draw [thick] (f.west) -- (e.east);
\draw [thick] (g.west) -- (f.east);

\node at (-0.5, 1) [label=left:$P_{1}$] (bot1) {};
\node at (-0.5, 0) [label=left:$P_{2}$] (bot2) {};

\draw [thick] (a.west) -- (bot1.east);
\draw [thick] (e.west) -- (bot2.east);
\draw [thick,->] (b) -- (f);
\end{tikzpicture}
\label{fig:vecclocks}
}
\subfloat[Slice]{
\begin{tikzpicture}
\node at (0,0) [place1] [label=above:{$[e]$}] (a) {};
\node at (0,0) [inner] {};

\node at (0,1) [place1] [label=above:{$[a]$}] (b) {};
\node at (0,1) [inner] {};

\node at (1.75,0) [place1] [label=120:{$[b,f]$}] (c) {};
\node at (1.75,0) [inner] {};

\node at (3.5,0) [place1] [label=above:{$[b,g]$}] (d) {};
\node at (3.5,0) [inner] {};

\node at (3.5,1) [place1] [label=above:{$[c,f]$}] (e) {};
\node at (3.5,1) [inner] {};

%\node at (1,-0.5) [place1] [label=right:$: \mbox{meta-event}$] (f) {};
%\node at (1,-0.5) [inner] {};

\draw [thick,->] (b) to [out=0,in=100] (c);
\draw [thick,->] (a) -- (c);
\draw [thick,->] (c) -- (d);
\draw [->,thick] (c) to [out=60,in=180] (e);
\end{tikzpicture}
%\caption{Computation }
\label{fig:slice}
}
\caption{A Computation, Vector Clock Representation, and Slice with respect to predicate $B = $``all channels are empty''}
\label{fig:computation}
\end{figure*}
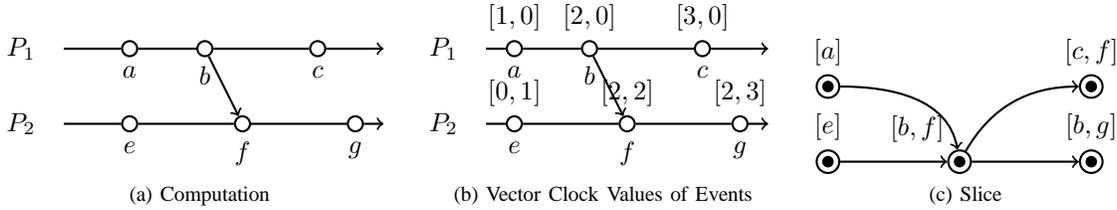

We now present some required concepts: 
\begin{defn}[Consistent Cut] 
Given a distributed computation $(E, \ra)$, a subset of events $C \subseteq E$ is said to form a
\emph{consistent cut} if $C$ contains an event $e$ only if it contains all events that happened-before $e$. Formally,
$
e \in C \land f \ra e \implies f \in C
$.
\end{defn}
\remove{
As an example, consider the computation shown in Fig~\ref{fig:computation}\subref{fig:comp}. The subset of 
events $\{a, b, e, i \}$ forms a consistent cut, whereas the subset $\{ a, b, 
e, f, i \}$ does not.}
\remove{because $j \ra f$ but $j$ does not belong to the subset.}

The concept of a consistent cut (or, a consistent global state) is identical to that of a down-set (or 
order-ideal) used in lattice theory \cite{DavPri:1990:CUP}. Intuitively, a 
consistent cut captures the notion of a global state of the system at some point during
its execution \cite{ChaLam:1985:TrCS}. 
\remove{
In this paper, both  
consistent cut and consistent global state may be used interchangeably. }

Consider the computation shown in Fig~\ref{fig:computation}\subref{fig:comp}. The subset of 
events $\{a, b, e\}$ forms a consistent cut, whereas the subset $\{a, 
e, f\}$ does not; 
because $b \ra f$ ($b$ happened-before $f$) but $b$ is
not included in the subset.
A consistent cut can also be represented with a vector clock notation. For any consistent cut $C$, its vector clock 
$C.V$ can be computed as $
C.V =$ component-wise-max$\{e.V~ | ~ event ~ e \in C\}$, where $e.V$ denotes the vector clock of event $e$. 
For this paper, we use a shortened notation
for a cut of the computation. A cut $C$ is denoted by 
the latest events on each process. Thus, \{$a, b, e$\}
is denoted by [$b,e$] and \{$a, e, f$\} is represented
as [$a, f$]. 

Table~\ref{tbl:cuts} shows 
all the consistent global states/cuts and their
corresponding vector clock values for the computation in 
Fig.~\ref{fig:computation}. \remove{A visual representatio 
of the lattice of consistent cuts for Fig.~\ref{fig:computation} can be found in Fig.~\ref{fig:lattice} in Appendix. } We now present additional notions
from lattice theory that are key to our approach.  
\begin{table}[b]
\caption{Consistent Global States of Fig.~\ref{fig:computation} and Predicate
Evaluation for B=``{\em all channels empty}''}
\centering
    \begin{tabular}{|l | c | c | c | c | }
        \hline
            \# & State & Cut Vec. Clock & Pred. Eval.\\ \hline
            $1$ & [] & $[0,0]$  & {\bf True}\\ \hline
            $2$ & [a] & $[1,0]$ & {\bf True}\\ \hline
            $3$ & [b] & $[2,0]$ & False\\ \hline
            $4$ & [c] & $[3,0]$ & False\\ \hline
            $5$ & [e] & $[0,1]$ & {\bf True}\\ \hline
            $6$ & [a,e] & $[1,1]$ & {\bf True}\\ \hline
            $7$ & [b,e] & $[2,1]$ & False\\ \hline
            $8$ & [b,f] & $[2,2]$ & {\bf True}\\ \hline
            $9$ & [b,g] & $[2,3]$ & {\bf True} \\ \hline
            $10$ & [c,e] & $[3,1]$ & False\\ \hline
            $11$ & [c,f] & $[3,2]$ & {\bf True}\\ \hline
            $12$ & [c,g] & $[3,3]$ & {\bf True}\\ \hline
\end{tabular}
\label{tbl:cuts}
\end{table}
\remove{
\begin{table}[b]
\caption{Consistent Global States of Fig.~\ref{fig:computation} and Predicate
Evaluation for B=``{\em all channels empty}''}
\centering
    \begin{tabular}{|l | c | c | c | c | }
        \hline
            \# & State & Cut Vec. Clock & Pred. Eval. & Join-Irred.\\ \hline
            $1$ & [] & $[0,0]$  & {\bf True} & No\\ \hline
            $2$ & [a] & $[1,0]$ & {\bf True} & Yes\\ \hline
            $3$ & [b] & $[2,0]$ & False & Yes\\ \hline
            $4$ & [c] & $[3,0]$ & False & Yes\\ \hline
            $5$ & [e] & $[0,1]$ & {\bf True} & Yes\\ \hline
            $6$ & [a,e] & $[1,1]$ & {\bf True} & No\\ \hline
            $7$ & [b,e] & $[2,1]$ & False & No\\ \hline
            $8$ & [b,f] & $[2,2]$ & {\bf True} & Yes\\ \hline
            $9$ & [b,g] & $[2,3]$ & {\bf True} & Yes \\ \hline
            $10$ & [c,e] & $[3,1]$ & False & No\\ \hline
            $11$ & [c,f] & $[3,2]$ & {\bf True} & Yes\\ \hline
            $12$ & [c,g] & $[3,3]$ & {\bf True} & No\\ \hline
\end{tabular}
\label{tbl:cuts}
\end{table}
}

\begin{defn}[Join]
\label{defjoin}
A join of a pair of global states is defined as 
the set-union of the set of events in the states. 
\end{defn}
\begin{defn}[Meet]
\label{defmeet}
A meet of a pair of global states is defined as 
the set-intersection of the set of events in the states. 
\end{defn}
For two global states $C_1$ and $C_2$, their join is 
denoted with $C_1 \join C_2$, whereas
$C_1 \meet C_2$ denotes their meet.

\remove{
Fig.~\ref{fig:slice}\subref{fig:sliceVec} presents the consistent cuts of the slice in Fig.~\ref{fig:slice}\subref{fig:sliceEvents}, 
translated into vector clock form. Observe that the cut $\{a,b,e,f,i,j\}$, when viewed in terms of its maximal events is the cut when $P_1$ has executed
$b$, $P_2$ has executed $f$, and by consistency requirement $P_3$ has executed $j$.}
\remove{(so that the send of the message at $j$ must happen 
before receive of that message at $f$)} 
%% Given two consistent cuts $G$ and $H$, we say that $G \leq H$ iff $G \subseteq H$. 
\remove{
From \cite{DavPri:1990:CUP}, we have that the set of all down-sets 
(order-ideals) forms a lattice under the $\subseteq$ relation. Using that 
result, we get:}
\begin{thm}
\cite{DavPri:1990:CUP,Mat:1989:WDAG}
Let $\cuts{E}$ denote the set of all consistent cuts of a computation $(E, \ra)$.
\remove{The set of consistent cuts of any distributed computation $(E, \ra)$}
$\cuts{E}$ forms a lattice under the relation $\subseteq$.
\end{thm}

%% We denote the set of consistent cuts of computation $(E, \ra)$ by $cuts{E}$. 
\remove{
For any lattice, we use $\join$ and $\meet$ to denote the \emph{join} and 
\emph{meet} operators, respectively. A lattice is \emph{distributive} if the join 
operator distributes over the meet operator \cite{DavPri:1990:CUP}. \remove{That is, $a \meet 
(b \join c) \equiv (a \meet b) \join (a \meet c)$.} 
For the lattice of consistent cuts, the join and meet operators are given by set-union ($\cup$) and set-intersection ($\cap$), respectively. 
Clearly, $\cup$ distributes overs $\cap$. Therefore, 
{\em 
the set of consistent cuts of any distributed computation $(E, \ra)$ forms a distributive lattice under the relation $\subseteq$.} 
}
A \emph{global predicate} (or simply a \emph{predicate}) is a boolean-valued 
function on variables of processes. Given a consistent cut, a predicate is evaluated 
on the state resulting after executing all events in the cut. A global 
predicate is \emph{local} if it depends only on variables of a single 
process. 
If a predicate $B$ evaluates to true for a consistent cut $C$, we 
say that ``$C$ satisfies $B$'' and denote it by $C_B$.\\
%As an example, consider the computation shown in Fig.~\ref{fig:computation} and the predicate $(3 \leq x_{1} \leq 4) \land (x_{2} < 3)$. The consistent cut $\{ a, e, i \}$ does not satisfy the predicate, whereas the consistent cut $\{ q, b, e, f, i, j \}$ does. 
\begin{defn}[Linearity Property of Predicates]
\label{defLin}
A predicate $B$ is said to have the linearity property, 
if for any consistent cut $C$, 
which does not satisfy predicate $B$, 
there exists a process $P_i$ such that a cut
that satisfies $B$ can never 
be reached from $C$ without advancing along $P_i$.\\
Predicates that have the linearity property are called {\bf 
linear predicates}. 
\end{defn}
For example, consider the cut $[b,e]$ of the computation shown in 
Fig.~\ref{fig:computation}\subref{fig:comp}. The cut 
does not satisfy the predicate ``all channels
are empty'', and for the given cut, progress must be made on $P_2$ to reach 
the cut $[b,f]$ which satisfies the predicate. 

The process $P_i$ in the above definition is called 
a {\em forbidden process}. For a computation involving $n$ processes, given a consistent cut 
  that 
does not satisfy the predicate, $P_i$
can be found in $O(n)$ time for most linear predicates
used in practice. 
\remove{
The optimal approach for 
 this search varies with the actual  
predicate.}
\\
To find a {\em forbidden process} given a consistent 
cut, a process first checks if the cuts needs to be 
advanced on itself; if not it checks the states in the
total order defined using process ids, and picks the first 
process whose state makes the predicate false on the cut.
\remove{
In this paper, given a set of all {\em forbidden processes} on a cut, on any process we follow the steps of Algorithm~
\ref{alg:forbid} to identify the index of the first 
process on which the cut must be advanced. 
\begin{algorithm}[!h]
\DontPrintSemicolon
\SetKwFunction{FindForbiddenIndex}{FindForbiddenIndex}
\SetCommentSty{small}
\SetProcNameSty{textbf}
\FindForbiddenIndex(Set $forbiddenProcs$, int $myprocid$)\;
\Indp \If{}{} 
}
The set of linear predicates has a subset, the 
set of {\em regular predicates}, that exhibits 
a stronger property.  

\begin{defn}[Regular Predicates]
A predicate is called {\em regular} if for any 
two consistent cuts $C$ and $D$ that satisfy the predicate, the consistent 
cuts given by $(C \meet D)$ (the meet of $C$ and $D$) and $(C \join D)$ (the 
join of $C$ and $D$) also satisfy the predicate. 
\end{defn}
\remove{Formally, predicate $B$ is 
regular if for all consistent cuts $C$ and $D$, 
\[
C_B \;\land\; D_B \implies\;\;
C \cap D_B \;\land\; C \cup D_B
\]
}
Examples of regular predicates include local predicates (\emph{e.g.}, $x \leq 4$), conjunction of local predicates (\emph{e.g.}, $(x \leq 4) \land (y \geq 2)$ where $x$ and $y$ are variables on different processes) and monotonic channel predicates (\emph{e.g.}, there are at most $k$ 
messages in transit from $P_{i}$ to $P_{j}$)  \cite{MitGar:2005:DC}. 
Table~\ref{tbl:cuts} indicates whether or not the predicate ``all channels empty'' is satisfied
by each of the consistent global cuts of the computation in 
Fig.~\ref{fig:computation}. To use {\em computation slicing} for 
detecting regular predicates, we first need to capture the notion of {\em join-irreducible 
elements} for the lattice of consistent cuts. 

\remove{
The set of consistent cuts of the computation that satisfy a regular predicate forms a \emph{sublattice} of the lattice of consistent cuts, and is therefore also a distributive 
lattice \cite{DavPri:1990:CUP,MitGar:2005:DC}.
}

\begin{comment}

\section{Background}
What are posets? 
What is the lattice of down-sets?
Join-irreducible elements
Birkhoff's theorem
Slicing

\end{comment}

%\subsection{Computation Slicing}

\begin{defn}[Join-Irreducible Element]
\label{def:jie} 
Let $L$ be a lattice. An element $x \in L$ is join-irreducible if
\begin{enumerate}
\item $x$ is not the smallest element of $L$
\item $\forall a,b \in L: (x = a \join b) \implies (x = a) \vee (x = b)$.
\end{enumerate}
\end{defn}
Intuitively, a join-irreducible element of a lattice is one that cannot be represented as the join of two distinct elements of the lattice, both different from itself.
For the lattice of consistent cuts of a distributed 
computation, the join-irreducible elements 
correspond to consistent cuts that can not be 
reached by joins (set-union) of two or more consistent
cuts. For the computation of Fig.~\ref{fig:computation}, the join-irreducible consistent cuts are:
$[a], [b], [c], [e], [b,f], [b,g]$. 
\remove{Fig.~\ref{fig:lattice}
 (in Appendix) shows the join-irreducible consistent 
 cuts of the sub-lattice induced by predicate ``all 
 channels empty'' for computation of Fig.~\ref{fig:computation}.} 
\subsection{Computation Slice}
A computation slice of a computation with respect to a predicate $B$ is a concise representation of 
all the consistent cuts of the computation that satisfy the predicate $B$. 
When the predicate $B$ is regular, the set of
consistent cuts satisfying $B$, $L_B$ forms a sublattice  
of $L$, that is the lattice of all consistent cuts 
of the computation $(E,\ra)$. $L_B$ can equivalently be represented
using its join-irreducible elements \cite{DavPri:1990:CUP}.
Intuitively, join-irreducible elements form the basis of the lattice. The 
lattice can be generated by taking 
 joins of its basis elements.
Let $J_B$ be the set of all join-irreducible elements of $L_B$.
Let $J_B(e)$ denote the least consistent cut that 
includes $e$ and satisfies $B$. Then,
it can be shown \cite{GarMit:2001:ICDCS} 
that
\[ J_B = \{ J_B( e) | e \in E \} \]
The $J_B(event)$ values, in vector clock notation, for each event 
of the computation in Fig.~\ref{fig:computation} are:\\
$J_B(a) = [1,0]$, 
$J_B(b) = [2,2]$, 
$J_B(c) = [3,2]$, 
$J_B(e) = [0,1]$, 
$J_B(f) = [2,2]$, 
$J_B(g) = [2,3]$. 
We can now define a computation slice formally.
\begin{defn}
Let $(E,\ra)$ be a computation and $B$ be a regular predicate. 
The slice of the computation with respect to
$B$ is defined as $(J_B, \subseteq)$.
\end{defn}

The definition given here is different from the one given in \cite{GarMit:2001:ICDCS} and
\cite{MitGar:2005:DC} but equivalent for regular predicates as shown in 
\cite{MitGar:2005:DC}.\\
{\bf Note}: It is important to observe that $J_B(e)$ does not necessarily exist. Also, 
multiple events may have the same $J_B(e)$.

\remove{
As an example, consider the computation shown in figure~\ref{fig:computation} and its slice shown in figure~\ref{fig:slice} with respect to the predicate $(3 \leq x_{1} \leq 4) \land (x_{2} < 3)$. The slice consists of three meta-events $\{ a, b, e, f, i ,j \}$, $\{ c, d \}$ and $\{ k \}$. It can be verified that every consistent of the computation that satisfies the predicate is a consistent cut of the slice, and vice versa.
The notion of a computation slice is based on Birkhoff's Representation Theorem (listed in the Appendix) for Finite Distributive Lattices \cite{DavPri:1990:CUP}. 
\remove{The theorem and its characterization
are briefly discussed in the Appendix. } 
The definition of slice presented above  is based on the one used in \cite{GarMit:2001:ICDCS} and is especially suited for regular predicates. An alternative definition of slice is based on directed graphs (possibly containing cycles) \cite{MitGar:2005:DC} and can handle more general predicates.
}

For the computation shown in Fig.~\ref{fig:computation}\subref{fig:comp}, 
\remove{
Table~\ref{tbl:cuts} lists (`In Slice' column) the consistent 
cuts that are included in the slice with respect to the 
predicate ``all channels empty'',} Fig.~\ref{fig:computation}\subref{fig:slice} presents a visual representation 
of the slice. 

\remove{
To compute the slice in an efficient manner, we use the {\em linearity property} \cite{garg02} defined in Definition~\ref{defLin}. 
Note that every regular predicate is linear, and thus the {\em linearity property} holds for all regular predicates. 
 It is possible to find a forbidden process 
in {\em O}($n$) time for many useful predicates, such as conjunctive predicate. We use this notion for efficiently computing the slice. 
}

A centralized online algorithm to compute $J_B$ was proposed in \cite{MitSen+:2007:TPDS}.  
In the online version of this centralized algorithm,  
 a pre-identified process, called {\em slicer} process, plays the role of the slice computing process. All the processes 
in the system send their event and local state values whenever their local states change. The {\em slicer}
process maintains a queue of events for each process in the system, and on receiving the data from a process 
adds the event to the relevant queue. In addition, the slicer process also keeps a map of events and 
corresponding local states for each process in the system. For each received event, the slicer appends the event and local state mapping to the respective map. For every event  $e$ it receives, the slicer computes $J_B(e)$ using the {\em linearity property}. 

The centralized approach suffers from the drawback of causing a heavy load of messages as well as computation 
on just one process, namely the {\em slicer} process. Thus, for any large distributed computation, this approach 
would not scale well. To address this issue, we propose a distributed algorithm that 
significantly reduces the computational, as well as the message load on any process. 

\section{A Distributed Online algorithm for Slicing}
\label{sec:distAlg}
\remove{
In this section we give a distributed online algorithm to compute
slice of a computation with respect to a regular predicate.
Let $(E, \ra)$ be a distributed computation and 
$L$ be its lattice of consistent global states. We are required
to compute the slice of $(E, \ra)$ with respect to a regular predicate
$B$. } 

In this section, we present the key ideas and 
routines for distributed online algorithm for computing the slice. The required optimizations that tackle 
the challenges listed in Section~\ref{sec:contri} 
are discussed later. 
\remove{Let there be $N$ processes $P_1, P_2,..P_N$ in the computation. } In our algorithm,
we have $n$ slicer processes, $S_1, S_2,...,S_n$,
one for every application process. All slicer processes
cooperate to compute the task of slicing $(E, \ra)$. Let $E$ be
partitioned into $n$ sets $E_i$ such that $E_i$
is the set of events that occurred in $P_i$.
In our algorithm, $S_i$ computes
\[ J_i(B) = \{ J_B( e) | e \in E_i \} \]
Observe that by the definition of join-irreducible 
consistent cut, $e \ra f$ implies $J_B(e) \subseteq J_B(f)$. 
Since all events in a process are totally ordered,
the set of consistent cuts generated by any $S_i$ are also totally
ordered.
%\hfill\\
%The job of the slicing process $S_i$ is to find J_B(e) for all e in its
%application process. Each slicing process will process
%at most $E$ events, possibly fewer.
%\remove{
{\bf Note}: In this paper, the symbol $\ra$ 
indicates a {\em happened-before} relation; whereas the symbol 
$\la$ in the pseudo-code denotes assignment operation.
%}

Algorithm~\ref{alg:distSlice} presents the distributed algorithm for online slicing with respect to a regular predicate $B$.  
Each slicer process has a token assigned to it that 
goes around in the system. Other slicer processes 
cooperate in maintaining and processing the token. 
The goal of the token for the slicer process $S_i$ is to compute $J_B(e)$ for
all events $e \in E_i$. Whenever the token has computed $J_B(e)$ it returns to 
its original process, reports $J_B(e)$ and starts computing $J_B(succ(e))$, $succ(e)$ being the immediate successor of event $e$.
The token $T_i$ carries with it the following data:
\begin{itemize}
\item
$pid$: Process id of the slicer process to which it belongs. 
\item
$event$: Details of event $e$, specifically the event id and event's vector 
clock, at $P_i$ for which this token is computing $J_B(e)$. The identifier 
for event $e$ is the 
tuple \textless$pid, eid$\textgreater ~ that identifies each event in the computation uniquely.
\item
$gcut$: The vector clock corresponding to the cut which is under consideration (a candidate for $J_B(e)$). 
\item
$depend$: Dependency vector for events in $gcut$. The dependency vector is updated each time the information of an event
is added to the token (steps explained later), and is used to decide whether or not some cut being considered 
is consistent. On any token, its vector $gcut$ is a consistent global state {\em iff}
for all $i$, $depend[i] \leq gcut[i]$.
\item
$gstate$: Vector representation of global state corresponding to vector $gcut$. It is sufficient to keep only
the states relevant to the predicate $B$.
\item
$eval$: Evaluation of $B$ on $gstate$. 
The evaluation is either true or false; in our 
notation we use the values: $\{predtrue, predfalse\}$.
\item $target$: A pointer to the unique event in the computation for which a token has to wait. The event need not 
belong to the local process. 
\remove{It is $inconsistent$ if $gcut$ is not a consistent cut. It is $predtrue$ if $gcut$ is consistent and
$gstate$ satisfies the predicate. It is $predfalse$, otherwise.}
\end{itemize}

%\newcommand{ei}[k]{e^{i}_k}
%Let $e\leftarrowe^1_i$ be the first event at $P_i$.
%Let $\ei{1}$ be the first event at $P_i$.
%Clearly, $J_B( \ei{1}) \supseteq J(\ei{1})$.
%The vector clock $e.V$ corresponds to $J(e)$.
%More generally, let $e$ be $k^{th}$ event ($k>1$) with
%$pred(e)$ as te predecessor event of $e$ on $P_i$.
%To compute $J_B( e)$, the token starts scanning from 
%$max( J(e), J_B( pred(e))$.
%If pred(e) does not exist, i.e., e is the first event on P_i, then
%the token starts form max( J(e)).

A token waits at a slicer process $P_i$ under three 
specific conditions:
\begin{itemize}
\item [(C1)] The token is for process $S_i$ and it has 
computed $J_B(pred(e))$, $pred(e)$ being the immediate
predecessor event of $e$, and is waiting for the arrival of $e$.
\item
[(C2)] The token is for process $S_i$ and it is computing $J_B( f)$, where $f$ is 
an event on $P_i$ prior to $e$. The computation of $J_B( f)$
requires the token to advance along process $P_i$.
\item
[(C3)] The token is for process $S_j$ such that  $j \neq i$, and it is 
computing $J_B( f)$ which requires the 
token to advance along process $P_i$.
\end{itemize}
\begin{algorithm}[t]
\DontPrintSemicolon
\SetKwFunction{ReceiveEvent}{ReceiveEvent}
\SetKwFunction{ReceiveToken}{ReceiveToken}
\SetKwFunction{EvaluateToken}{EvaluateToken}
\SetKwFunction{AddEventToToken}{AddEventToToken}
\SetKwFunction{ProcessToken}{ProcessToken}
\SetKwFunction{SendIfNeeded}{SendIfNeeded}
\SetKwFunction{ReceiveStopSignal}{ReceiveStopSignal}
\SetCommentSty{small}
\SetProcNameSty{textbf}
\ReceiveEvent(Event $e$,State $localstate_e$)\nllabel{linelocalProgressStart}\;
\Indp save \textless$e.eid,localstate_e$\textgreater ~in local state map \textless$procstates$\textgreater\;\nllabel{savetolocalstate}
\ForEach{waiting token t at $S_i$}{
\Indp \uIf(//$t$ waiting for event $e$){$(t.target = e)$}{\nllabel{addtargetstart}
\AddEventToToken($t$,$e$)\nllabel{lineaddeventtotoken}\;
\ProcessToken($t$)\;
}\nllabel{addtargetend}
}\nllabel{linelocalProgressEnd}
\BlankLine
\Indm \AddEventToToken(Token $t$,Event $e$)\;\nllabel{addeventstart}
\Indp       $t.gstate[e.pid] \leftarrow procState[e.eid]$\;
           $t.gcut[e.pid] \leftarrow e.eid$\;
           \If(//my token: update token's $event$ pointer){$(t.pid = i)$}{
               $t.event = e$\;
           }
           $t.depend \leftarrow max(t.depend, e.V)$ //
           set causal dependency\nllabel{updateDepend}\;\nllabel{addeventend}

\BlankLine
\Indm \ProcessToken(Token $t$)\;\nllabel{proceventstart}
\Indp \uIf{($t.gcut$ is $inconsistent$)}{\nllabel{proceventconsistentcheck} %< t.depend
\Indm \tcc*[l]{find $k$ : $t.gcut[k] < t.depend[k]$}
 \Indp \Indp     $t.target \la t.gcut[k] + 1$ //~set desired event\;
      {\bf send} $t$ to $S_k$\;\nllabel{sendincosistent}
   }
   \Else(// $t.gcut$ is consistent){
   \EvaluateToken($t$)\;
   }\nllabel{proceventend}
\BlankLine
\Indm
{\bf \EvaluateToken}(Token $t$)\;\nllabel{evaltokenstart}
\Indp     \uIf(//$B$ is true on cut given by $t.gcut$){$B(t.gstate)$}{\nllabel{BTrueOnCut}
 %                 \UpdateStalledTokens($t$)\;
                  $t.eval \leftarrow predtrue$\;
                  {\bf send} $t$ to process $S_{t.pid}$\;\nllabel{linesendTokenToParent}
              }
              \Else(// $B$ is false on $t.gstate$){
                 $t.eval \leftarrow predfalse$\;
\Indm \Indm \tcc*[l]{$P_k$ : forbidden process in $t.gstate$ for $B$}
\Indp \Indp $t.target \leftarrow t.gcut[k]+1$\;\nllabel{linesettargetforfalse}
                  {\bf send} $t$ to $S_k$\;\nllabel{linesendtokentopk}
            }\nllabel{evaltokenend}
\BlankLine
\Indm \ReceiveToken(Token $t$)\;\nllabel{recvtokenstart}
\Indp \uIf(//my token, $B$ true){$(t.eval = predtrue) \land (t.pid = i)$}{
       %        $localJBes.add$$<t.eid,t.gcut>$\;
               {\bf output}$(t.pid, t.eid, t.gcut)$\;\nllabel{recvOutput}
\Indm \Indm              \tcc*[l]{token waits for the next event}
\Indp \Indp $t.target \leftarrow t.gcut[i]+1$\;\nllabel{settargetnextevent}
            $t.waiting \leftarrow true$\;\nllabel{setwaitfornext}
}
\Else(//either incosistent cut, or predicate false){
        $newid \leftarrow t.target$ //~id of event $t$ requires\;
\If{($\exists f \in localEvents : f.id = newid$)}{\nllabel{lookforreqevent}
//required event  has happened  \;
         \AddEventToToken($t$,$f$)\;\nllabel{leastevent}
         \EvaluateToken($t$)\;\nllabel{evaltokenafterleast}
        }
        //else, the token remains in $waiting$ state
}\nllabel{recvtokenend}
\BlankLine
\Indm \ReceiveStopSignal\;\nllabel{stopsignalstart}
\Indp \ForEach{token t : t.pid $\neq$ i}{
   //not my token, send back to parent\;
   {\bf send} t to $S_{t.pid}$\;
   }\nllabel{stopsignalend}
\caption{Algorithm at $S_i$}
\label{alg:distSlice}
\end{algorithm}
On occurrence of each relevant event $e \in E_i$,  
the computation process $P_i$ performs a {\em local} 
enqueue to {\em slicer} $S_i$, with the details of this event. Note that $P_i$ and its slicer $S_i$ are modeled as 
two threads on the same process, and therefore the {\em local} enqueue is simply an insertion 
into the queue (that is shared between the threads on the same process) of the 
{\em slicer}. 
The message contains the details of event $e$, i.e. the event identifier \textless$eid,pid$\textgreater \hspace{1pt}, the corresponding vector clock $e.V$, and $P_i$'s local state $localstate_e$ corresponding 
to $e$. The steps of the presented routines are explained 
below:\\
\hfill\\
{\tt ReceiveEvent} (Lines~\ref{linelocalProgressStart}-\ref{linelocalProgressEnd}): 
On receiving the details of event $e$ from $P_i$, 
$S_i$ adds them in the mapping of $P_i$'s local states $procstates$ (line~\ref{savetolocalstate}).
It then iterates over all the waiting tokens, and 
checks their $target$. For each token that has
$e$ as the target (required event to make progress), 
$S_i$ updates the state of the token, and then 
processes it.\\
\hfill\\
{\tt AddEventToToken} (Lines~\ref{addeventstart}-\ref{addeventend}): 
To update the state of some token $t$ on $S_i$, we advance the candidate cut to include the new event
by setting $t.gcut[i]$ to the id of event $e$. If 
$S_i$ is the parent process of the token ($T_i$), 
then the $t.event$ pointer is updated to indicate 
the event id for which token is computing the join-irreducible cut that satisfies the predicate. The
causal-dependency is updated at line~\ref{updateDepend}, 
which is required for checking whether or not the cut 
is consistent. \\
\hfill\\
%\vspace{-1pt}
{\tt ProcessToken} (Lines~\ref{proceventstart}-\ref{proceventend}): 
To process any token, $S_i$ first checks that the global state in the token is 
 consistent (line~\ref{proceventconsistentcheck}) and at least beyond the global states
that were earlier evaluated to be false.
For $t$'s evaluation of a global cut $t.gcut$ to be consistent, $t.gcut$ must be at least $t.depend$. This is verified by 
checking the component-wise values in both these vectors.
If some index $k$ is found where $t.depend > t.gcut$, the 
token's cut is inconsistent, and $t.gcut$ must be advanced
by at least one event on $P_k$, by sending the 
token to {\em slicer} of $P_k$. 
If the cut is consistent, the predicate is evaluated on the variables stored as part of $t.gstate$ by calling the {\tt EvaluateToken} routine. \\
\hfill\\
{\tt EvaluateToken} (Lines~\ref{evaltokenstart}-\ref{evaltokenend}): 
The cut represented by $t.gstate$ is evaluated;
if the predicate is true, then the token has 
computed $J_B( e)$ for the event
$e=$\textless$t.pid,t.eid$\textgreater. The token 
is then sent to its parent {\em slicer}. 
If the evaluation of the predicate on the cut is false, the $target$ pointer 
is updated, at line \ref{linesettargetforfalse}, and the token is sent to the {\em forbidden} process on which the 
token must make progress.\\
\hfill\\
{\tt ReceiveToken} (Lines~\ref{recvtokenstart}-\ref{recvtokenend}): On receiving a token, the {\em slicer} checks if the predicate evaluation on the token is 
true, and the token is owned by the {\em slicer}. In such a case, 
the {\em 
slicer} outputs the cut information, and now uses the
token to find $J_B( succ(e))$, where $succ(e)$ 
denotes the event that locally succeeds $e$. This is done by setting the new event id in $t.target$ at line 
\ref{settargetnextevent}, and then setting the waiting
flag (line~\ref{setwaitfornext}).
If the predicate evaluation on the token is false, then 
the $target$ pointer of the token points to the 
event required by the token to make progress. $S_i$
looks for such an event (line~\ref{lookforreqevent}), and if it has been reported
to $S_i$ by $P_i$, then adds that event (line~\ref{leastevent}) to the token and 
processes it (line~\ref{evaltokenafterleast}).  
In case the desired event has not been reported yet 
to the {\em slicer}
process, the token is retained at the process $S_i$ and is kept in the waiting state until the required
event arrives. Upon arrival of the required event, its details are added to the token and the token is processed. \\
{\bf Note}: The notation of $target \la t.gcut[i] +1$ 
means that if the $t.gcut[i]$ holds the event id \textless$pid,eid$\textgreater, then the $target$ pointer is set to 
\textless$pid,eid+1$\textgreater. \\
\hfill\\
{\tt ReceiveStopSignal} (Lines~\ref{stopsignalstart}-\ref{stopsignalend}): For finite computations, a 
single token based termination detection algorithm is 
used in tandem. Any one of the {\em slicer} process, let us assume $S_1$ holds a separate {\em stop}
token. Whenever $P_1$ is finished with 
its computation, it sends a signal to $S_1$, and $S_1$ in turn checks
if it has any tokens on which it has not updated the local events from $P_1$.
Only after all such updates and processing is completed, and there are no 
more local events to process, $S_1$ forwards the {\em stop} 
token to $S_2$, and so on. When $S_1$  receives 
the {\em stop} token from $S_n$, it can deduce that all the {\em slicer} processes
have completed processing all the events from their local queues, and there
is no {\em slicing} token that can be advanced further.   
$S_1$ then sends the `stop' signal to all the {\em slicer} processes, 
including itself. On receiving the `stop' signal, $S_i$ sends all the {\em slicing} tokens 
that do not belong to it
 back to their parent processes. 
%\remove{

Note that the routines require 
atomic updates and reads on the local queues, as well as
on tokens 
present at $S_i$. In the interest of space 
we skip presenting the lower level implementation details,
that involve common local synchronization techniques. 
%}
\subsection{Example of Algorithm Execution}
\label{sec:run_ex}
This example illustrates the algorithm execution steps 
for one possible run (real time observations) of 
the computation shown in Fig.~\ref{fig:computation}, 
with respect to the predicate
$B=$ ``all channels empty''. 
The algorithm starts with two slicing processes $S_1$ and $S_2$, each having a token -- $T_1$ and $T_2$ respectively. 
The {\em target} pointer for each token $T_i$ is initialized  to
the event \textless$i,1$\textgreater. When  
event $a$ is reported, $S_1$ adds its details to $T_1$, and 
on its evaluation finds the predicate ``all channels empty'' to be true, and outputs this information. It 
then updates $T_1.target$ pointer and waits for the 
next event to arrive. Similar steps 
are performed by $S_2$ on $T_2$ when $e$ is reported.

When $b$ is 
reported to $S_1$, and $T_1$ is evaluated with the updated
information, the predicate is false on the state $[b]$.
Given that $b$ is a message send event, it is obvious 
that for the channel to be empty, the message receive 
event should also be incorporated. Thus, $S_1$ sends
$T_1$ to $S_2$ after setting the target pointer to 
the first event on $S_2$. On receiving $T_1$, $S_2$ 
fetches the information of its first event ($e$) and 
updates $T_1$. The subsequent evaluation still 
leads to the predicate being false. Thus $S_2$ 
retains $T_1$ and waits for the next event. 

When $f$ 
is reported, $S_2$ updates both $T_1$ and $T_2$ with $f$'s 
details. $S_2$'s evaluation on $T_1.gstate$, represented
by $[b,f]$ is true, 
and as per line \ref{linesendTokenToParent}, $T_1$ is 
sent back to $S_1$ where the consistent cut $[b,f]$
is output. $T_1$ now waits for the next event.
However, after being updated with the details of event 
$f$, the resulting cut on $T_2$
is inconsistent, as the message-receive information 
is present but the information regarding the 
corresponding send event is missing. By using the 
vector clock values, $T_2$'s 
target would be set to the id of message-send event $b$.
$S_2$ would then send $T_2$ to $S_1$. On receiving $T_2$, 
$S_1$ finds the required event (looking at $T_2.target$)
and after updating $T_2$ with its details, evaluates 
the token. The predicate is true on $T_2.gstate$ now, 
and $T_2$ is sent back to $S_2$. On receiving $T_2$, 
$S_2$ outputs the consistent cut $[b,f]$, and 
waits for the next event. On receiving details of 
event $c$, and adding them to the waiting token $T_1$, 
the predicate is found to be true again on $T_1$, and 
$S_1$ outputs $[c,f]$. Similarly on receiving $g$, $S_2$
performs similar steps and outputs $[b,g]$. 
Note that the consistent cuts $[a,b]$ and $[c,g]$, 
both of which satisfy the predicate are not enumerated
as they are not join-irreducible, and can be
constructed by the unions of $[a]$, $[b]$ and 
$[c,f]$, $[b,g]$ respectively. 
\hfill\\
\remove{
\subsection{Correctness} 
In the interest of space,  the detailed proof of correctness of the algorithm has been shifted to 
the Appendix.
}
%~\ref{sec:appendix}. 

\subsection{Proof of Correctness} 
%\subsection{Proof of Correctness for Algorithm~\ref{alg:distSlice}}
%\newcommand{\pf}{IEEEproof}
\newcommand{\jbei}{$J_B(e)$ }
\newcommand{\jbe}{$J_B(e)$ }
\newcommand{\einei}{$e \in E_i$ }
\newcommand{\eine}{$e \in E$ }
\newcommand{\tigcut}{$T_{i}.gcut$ }
This section proves correctness and termination (for finite computations) of the distributed algorithm of Algorithm~\ref{alg:distSlice}. The proofs 
presented here are for finite computations. The correctness argument can be easily extended to infinite computations. 

\begin{lem}\label{nodeadlockorig} The algorithm presented in Algorithm~\ref{alg:distSlice} does not deadlock. 
\end{lem}
\begin{\pf}
 The algorithm involves $n$ tokens, and 
none of the tokens wait for any other token to complete any task. With 
non-lossy channels, and no failing processes, the tokens are never lost. The 
progress of any token depends on the $target$ event, and as per lines
~\ref{addtargetstart}-\ref{addtargetend}, whenever an event is reported to a {\em slicer}, 
it always updates the tokens with their $target$ being this event. Thus, 
the algorithm can not lead to deadlocks. 
\end{\pf}

\begin{lem}\label{lemAdvance} If a token $T_i$ is evaluating $J_B(e)$ for $e \in E_i$, assuming $J_B(e)$ exists, and if $T_i$.$gcut < J_B(e)$, then
$T_{i}.gcut$ would be advanced in finite time.\end{lem}
\begin{\pf}
If during the computation of $J_B(e)$, at any instance $T_i$.$gcut < J_B(e)$, then there are two possibilities 
for $gcut$:\\
(a) $gcut$ is consistent: This means that the evaluation of predicate $B$ on $gcut$ must be false, as by definition 
\jbei 
is the least consistent cut that satisfies $B$ and includes $e$. In this case, by line
\ref{linesettargetforfalse} and subsequent steps, the token would be forced to advance on some process.\\ 
(b) $gcut$ is inconsistent: The token is advanced on some process by execution of lines~
\ref{proceventconsistentcheck}-\ref{sendincosistent}. 
\end{\pf}
\begin{lem}\label{jbeoutput} While evaluating $J_B(e)$ for event $e \in E_i$ on 
token $T_i$, if $T_i$.$gcut < J_B(e)$ currently and $J_B(e)$ exists then the algorithm eventually outputs $J_B(e)$.\end{lem}
\begin{\pf}
By Lemma \ref{lemAdvance}, the global cut of $T_i$ would be advanced in finite time. Given that $J_B(e)$ exists, we know that by the linearity property, there
must exist a process on which $T_i$ should progress its $gcut$ and $gstate$ 
vectors in order to reach the $J_B(e)$;
lines~\ref{linesettargetforfalse}-\ref{evaltokenend} ensure 
that this forbidden process is found and $T_i$ sent to this process. By the previous Lemma, the cut 
on the $T_i$ would be advanced until it matches $J_B(e)$.
By line \ref{recvOutput} of the algorithm, whenever $J_B(e)$ is reached, it would be output.  
\remove{If for event $e \in E_i$, $J_B(e)$ never exists, then the result is trivial as no global cut
having $e$ 
would be output and the token $T_i$
would be returned to $S_i$ at the end of the algorithm. 
Observe that initially $T_{i}.gcut < J_B(e)$ as the $gcut$ vectors of all the tokens are empty 
Given the linear progress of the algorithm, by 
As every token is evaluated on each event update, $T_i$ would be evaluated by some process whenever 
its $gcut$ matches $J_B(e)$. If this evaluation is carried out by $S_i$, then 
by line \ref{lineOutput}, the algorithm would output $J_B(e)$. Otherwise if the evaluation 
was performed by a process other than $S_i$, then by line \ref{linesendTokenToParent} 
it would send $T_i$ back to $S_i$, who would output \jbei as per line \ref{recvOutput}
}
\end{\pf}
\begin{lem}\label{noadvancebeyond} For any token $T_i$, the 
algorithm never advances $T_{i}.gcut$ vector beyond $J_B(e)$ on any process, when searching $J_B(e)$ for $e \in E_i$. 
\end{lem}
\begin{\pf}
The search for $J_B(e)$ starts with either an empty global state vector, or from the global state that is at least 
$J_B(pred(e))$, where $pred(e)$ is the immediate
predecessor event of $e$ on $S_i$. Thus, till $J_B(e)$ is reached, the global cut under consideration is always less than $J_B(e)$.
From the linearity property of advancing on the {\rm forbidden process}, and Lemma \ref{lemAdvance}, the cut would 
be advanced in finite time. Whenever the cut reaches $J_B(e)$, it would be output as per Lemma \ref{jbeoutput} and 
the token would be sent back to its parent slicer, to either begin the search for $succ(e)$ or to wait for $succ(e)$ to arrive ($succ(e)$ being the immediate 
successor of $e$).  
Thus, $T_{i}.gcut$ would never advance beyond $J_B(e)$ on any process when searching for $J_B(e)$ for any event $e$. 
\end{\pf}
\remove{
Initially, for any token the $gcut$ vectors are empty. When an event $e$ arrives 
on $S_i$, by line \ref{lineaddeventtotoken} of the algorithm 
$S_i$ would add it to all the waiting tokens it holds. If $S_i$ currently holds $T_i$
then $e$'s details would be added to $T_i$. After this update $T_i$ would be evaluated, and if the predicate $B$ is satisfied, then 
$T_{i}.gcut = J_B(e)$ holds. At this stage, as per line \ref{lineWaitForNextEvent}, the search for 
of \jbei on $T_i$ halts, and thus does never advances its $gcut$ beyond \jbei
within the computation. However if the evaluation indicates that either the predicate 
is false, or the $gcut$ is inconsistent, then $T_i$ would be sent to some 
corresponding forbidden process to resolve this. Whichever process $T_i$ is sent to, 
that process would also follow the steps of updating the token and evaluating it.
And by inductive reasoning for any such process, we can argue as above that 
the $gcut$ vector of $T_i$ during the search for \jbei would never be advanced beyond
$J_B(e)$. 
\end{\pf}
}
\begin{lem}\label{lemTokenReturn} If token $T_i$ is currently not at $S_i$, then $T_i$ would return to $S_i$ in finite time. 
\end{lem} 
\begin{\pf} Assume $T_i$ is currently at $S_j$ ($j \neq i$). 
$S_j$ would advance \tigcut in finite time as per Lemma 
\ref{lemAdvance}. With no deadlocks (Lemma~\ref{nodeadlockorig}), and
by the results of Lemma \ref{jbeoutput} and Theorem \ref{noadvancebeyond}, we are guaranteed that if $J_B(T_i.event)$ exists then within a finite time, \tigcut
vector would be advanced to $J_B(T_i.event)$ and $T_i$ would be sent back to $S_i$.
If $J_B(T_i.event)$ does not exist then at least one slicer process $S_k$ would run out of 
all its events while attempting to advance on \tigcut. In such a case, 
knowing that there are no more events to process, $S_k$ would send 
$T_i$ back to $S_i$ (lines~\ref{stopsignalstart}-\ref{stopsignalend}).
\end{\pf}
\begin{thm}\label{finiteEi} {\bf (Termination)}: For a 
 finite computation, the algorithm terminates in finite time.\end{thm} 
\begin{\pf}
We first prove that  for any event $e \in E_i$, computation
of finding $J_B(e)$ with token $T_i$ takes finite time.
By Lemma \ref{lemAdvance}, $T_i$ always advances in finite time while computing $J_B(e)$. If \jbei exists, then based on this 
observation 
within a finite time the token $T_i$ would advance its $gcut$ to $J_B(e)$, if it exists. By Lemma \ref{jbeoutput}, the algorithm 
would output this cut, thus finishing the \jbei search and as per Theorem \ref{noadvancebeyond} would not advance 
any further for \jbei computation. Thus, if \jbei exists then it would be output in finite time. By Lemma \ref{lemTokenReturn} 
the token would be returned to its parent process and the \jbei computation for \einei would finish in finite time.  

If \jbei does not exist, then as we argued in Lemma \ref{lemTokenReturn} some {\em slicer} would run out of events to process in the 
finite computation, and thus return the token to $S_i$, which would result in search for \jbei computation to terminate. As each of these
steps is also guaranteed to finish in finite time as per above Lemmas, we conclude that \jbei computation for \einei 
finished in finite time. 

Now we can apply this result to all the events in $E$, and guarantee termination in finite time. 
\end{\pf}
\begin{thm}\label{correctness}The algorithm outputs all the elements
of $J_B$.  
\end{thm}
\begin{\pf}
Whenever any event $e \in E$ occurs, it is reported by some process $P_i$ on which 
it occurs, to the corresponding slicer process $S_i$. Thus $e$ can be represented 
as \einei. If at the time $e$ is reported to $S_i$, $T_i$ is held by $S_i$ then 
by Lemmas \ref{lemAdvance} and \ref{jbeoutput}, it is guaranteed that the algorithm 
would output $J_B(e)$. If $S_i$ does not hold the token $T_i$ when $e$ is reported to it, 
then by Lemma \ref{lemTokenReturn}, $T_i$ would arrive on $S_i$ within finite time. If $S_i$ has any other 
events in its processing queue before $e$, then as per Theorem \ref{finiteEi}, $S_i$ would finish those 
computations in finite time too. Thus, within a finite time, the  
computation for finding \jbei with $T_i$ would eventually be started by $S_i$. Once
this computation is started, the results of Lemmas \ref{lemAdvance} and \ref{jbeoutput}
can be applied again to guarantee that the algorithm would output $J_B(e)$, if it exists. 

Repeatedly applying this result to all the events in $E$, we are guaranteed that 
the algorithm would output \jbe for every event \eine. Thus the algorithm outputs all 
the join-irreducible elements of the computation, which by definition together
form $J_B$. \end{\pf}
%\end{IEEEproof}
\remove{
\begin{lem}\label{lemProgress}{\bf (Termination)}: The algorithm terminates in finite time.  \end{lem} 
\begin{IEEEproof}
The result follows by applying Theorem \ref{finiteEi} to all the events in $E$. 
\end{IEEEproof}
}
\begin{thm}\label{noJunk} The algorithm only outputs join-irreducible global states that satisfy predicate $B$.  \end{thm} 
\begin{IEEEproof}
By Lemma \ref{noadvancebeyond}, while performing computations for \einei on token $T_i$, the algorithm would not advance on token 
$T_i$ beyond $J_B(e)$. Since only token $T_i$ is responsible for computing \jbei for all the events \einei, the algorithm 
would not advance beyond \jbei on any token. In order to output a global state that is not  join-irreducible we must advance the cut of 
at least one token beyond a least global state that satisfies $B$. The result follows from the above assertions.  
\end{IEEEproof}
\remove{\begin{thm}{\bf (Correctness)}: The distributed online slicing algorithm \remove{presented in Algorithm \ref{alg:distSlice}} is correct and terminates in finite time.\end{thm} 
}

Theorem \ref{finiteEi} guarantees termination, and  correctness follows from Theorems \ref{correctness},  and \ref{noJunk}. 

\label{sec:opts}
\section{Optimizations}
The distributed algorithm presented in the previous section is not optimized to avoid redundant token messages, as well 
as duplicate computations. 
Whenever a {\em slicer} process $S_i$ needs to send any token 
to another process $S_k$, it should first check if it 
currently holds the token $T_k$, and if the desired
information is present in $T_k$. If the information 
is available, the token $T_i$ can be updated with the information without being 
sent to $S_k$; and only if the details of required event are not 
available locally, the token is sent to $S_k$. These steps
are captured in the procedure {\tt SendIfNeeded} 
shown in Algorithm~\ref{alg:checksend}. 
\begin{algorithm}[]
\DontPrintSemicolon
\SetKwFunction{SendIfNeeded}{SendIfNeeded}
\SetCommentSty{small}
\SetProcNameSty{textbf}
\SendIfNeeded(Token $t$, int $k$)\;
\tcc*[l]{$k$: id of the slicer process to which $t$ should be sent}
\Indp
 \While{($k \neq i) \land$ (have $token_k$)}{
\Indm \Indm  \tcc*[l]{$t$ should be sent to $S_k$, and $S_i$ has $S_k$'s token}
\Indp
\uIf(//$token_k$ has info of $t$'s //target event){($t.target = token_k.event$)}{
      $t.gcut[k] \la token_{k}.gcut[k]$\; %
      $t.depend[k] \la token_{k}.depend[k]$\; %
      $t.state[k] \la token_{k}.state[k]$\; %
      \uIf(//still inconsistent){($t.gcut$ is inconsistent)}{
\Indm     \tcc*[l]{find $j$ : $t.gcut[j] < t.depend[j]$}
     \Indp \Indp $t.target \la t.gcut[j] + 1$\;
               $k \la j$ ~//set $k$ for while condition\;
      }
      \Else(//~$t.gcut$ is consistent now, evaluate){
         \EvaluateToken($t$)\;      
      }
}\Indp
\Else(//~desired event details not in $token_k$){
   $break$\;
}
}
\Indm     \tcc*[l]{desired token or event info not present}
\Indp
\If{($t.target.pid \neq i$)}{
\Indm     \tcc*[l]{target event on some other process}
       {\bf send} $t$ to $S_k$\; 
       }
       
       %}
       \remove{
       \Else(\tcc*[l]{$gcut$ inconsistent; find $j$ for which $t.gcut[j] < t.depend[j]$}) { 
       \uIf{$(j = t.pid)$}{
          send $t$ to $S_j$ \nllabel{linesendTokenToSj}\;
          }
      \Else{
%           wait for $S_j$ to instruct\;  
            $t.stalled \leftarrow true$\;
          }
    }
    }
\caption{SendIfNeeded at $S_i$}
\label{alg:checksend}
\end{algorithm}

There are additional optimizations that significantly reduce the number of token messages. It is easy to 
observe that in the proposed form of Algorithm~\ref{alg:distSlice}, the algorithm performs many 
redundant computations. This redundancy is caused by computations of $J_B(e)$ and $J_B(f)$ where $e \neq f$, and 
$J_B(e) = J_B(f)$. In this case, given that both the join-irreducible consistent cuts are same, it would suffice that 
the algorithm only compute either of them. For this purpose, we first 
present some additional results:\\

%Let us consider the computation of Fig.~\ref{fig:computation} for the
%predicate $B = (x_1 > 1) \wedge (x_2 > 1)$. 

\begin{lem}
\label{lem:subsetJB}
$f \in J_B(e) \Rightarrow J_B(f) \subseteq J_B(e)$. 
\end{lem}
\begin{\pf}
$J_B(f)$ is the least consistent cut of the computation that satisfies the predicate, and contains $f$.
$J_B(e)$ includes $f$, and satisfies the predicate. Therefore $J_B(f) \subseteq 
J_B(e)$.
\end{\pf}

\begin{lem}
\label{lem:bothinJB}
$f \in J_B(e) \wedge e \in J_B(f) \Rightarrow J_B(e) = J_B(f)$. 
\end{lem}
\begin{\pf}
Apply previous Lemma twice.  \end{\pf} 
\begin{lem}
\label{lem:happBefJB}
$e \ra f \wedge f \in J_B(e) \Rightarrow J_B(f) = J_B(e)$. 
\end{lem}
\begin{\pf}
By Lemma~\ref{lem:subsetJB}, $f \in J_B(e)$ implies that $J_B(f) \subseteq J_B(e)$ must hold. 
 Given $e \ra f$, 
by the consistency requirement $J_B(f)$ must contain $e$. Thus, $J_B(e) \subseteq J_B(f)$. 
\remove {is the least consistent cut that satisfies 
the predicate and contains $f$; however it must also contain $e$. Thus, $J_B(f)$ should be the least consistent 
cut that satisfies the predicate and contains both $e$ and $f$. Given $f \in J_B(e)$, $J_B(e)$ is the least 
consistent cut that contains both $e$ and $f$ and satisfies the predicate. Hence, $J_B(f) = J_B(e)$.  }
\end{\pf}
\hfill\\
In order to prevent computations that result in identical join-irreducible states, we modify the proposed distributed algorithm 
of Algorithm~\ref{alg:distSlice} to incorporate Lemmas~\ref{lem:bothinJB}, and \ref{lem:happBefJB}. The modified algorithm is 
presented in Algorithm~\ref{alg:optimized}.
\begin{algorithm}[!ht]
\DontPrintSemicolon
\SetKwFunction{ReceiveEvent}{ReceiveEvent}
\SetKwFunction{ReceiveToken}{ReceiveToken}
\SetKwFunction{EvaluateToken}{EvaluateToken}
\SetKwFunction{AddEventToToken}{AddEventToToken}
\SetKwFunction{UpdateStalledTokens}{UpdateStalledTokens}
\SetKwFunction{ReceiveCut}{ReceiveCut}
\SetKwFunction{CopyCutIfNeeded}{CopyCutIfNeeded}
\SetKwFunction{ProcessToken}{ProcessToken}
\SetCommentSty{small}
\SetProcNameSty{textbf}
%\KwIn{$P_i$ sends event $e$ to slicer process $S_i$}
%\KwData{$procStates$}
%\KwOut{All join-irreducible elements of $L_B$}
%\BlankLine
\remove{
\ReceiveEvent(Event $e$,State $localState$)\nllabel{optlinelocalProgressStart}\;
\Indp save \textless $e.eid,localState$ \textgreater in local state queue\;
\ForEach{waiting token t at $S_i$} {
\Indp \If{$(t.target = e)$}{
\AddEventToToken($t$,$e$)\nllabel{optlineaddeventtotoken}\;
\ProcessToken($t$)\;
}
}\nllabel{optlinelocalProgressEnd}
\BlankLine
}
\AddEventToToken(Token $t$,Event $e$)\;
\Indp       $t.gstate[e.pid] \leftarrow procState[e.eid]$\;
           $t.gcut[e.pid] \leftarrow e.eid$\;
           \If(//my token: update current event){$(t.pid = i)$}{
               $currentE = e.eid$\; 
               $t.event = e$\;
           }
           $t.depend \leftarrow max(t.depend, e.V)$\nllabel{optupdateDepend}\;
\BlankLine
\Indm \ProcessToken(Token $t$)\;
\Indp \If{($t.gcut$ is $inconsistent$)}{ %< t.depend
\If(//~message receive event){($t.event.type = MSGRECV$)}{ \nllabel{recvWaitStart}
      $t.target \la t.event.sender$\; %t.depend[k]$
   \If(//~my token){$(t.pid=i)$}{
      $t.stalled \la true$ //stall, wait to hear from target\;
%      $mytarget \la t.target$\;
   }\nllabel{recvWaitEnd}
   \Else(//~progress other processes' token){
\Indm \Indm \tcc*[l]{find $k$ s.t. $t.gcut[k] < t.depend[k]$}
\Indp
      $t.target \la t.depend[k]$\; %
                  \SendIfNeeded($t$) //~use optimized send approach\;
%      send $t$ to $S_k$\; 
   }
}
}
   \Else(// $t.gcut$ is consistent, evaluate its state){
   \EvaluateToken($t$)\;
   }

\Indm \ReceiveToken(Token $t$)\;
\Indp \If(//~my token, $B$ true){$(t.eval = predtrue) \land (t.pid = i)$}{
       %        $localJBes.add$$<t.eid,t.gcut>$\;
               {\bf output}$(t.pid, t.eid, t.gcut)$\nllabel{optrecvOutput}\;
               \UpdateStalledTokens($t.eid$,$t.gcut$)\;\nllabel{updateStalledCall}
\Indm \Indm              \tcc*[l]{token waits for the next event}
\Indp \Indp $t.target \leftarrow t.gcut[i]+1$\;
            $t.waiting \leftarrow true$\;
}
\Else(//either incosistent cut, or predicate false){
        $newid \leftarrow t.target$ ~~~~~//~id of $t$'s required event\;
	\If(//~no causal-dependency, and symmetry breaking){$t.event \not\ra currentE \land newid \geq currentE \land i > t.pid$}{\nllabel{concStallStart}
	   $t.stalled \leftarrow true$ ~~~~~//stall the token\;
%	   {\bf return}\;
	}\nllabel{concStallEnd}
	\ElseIf{($\exists f \in localEvents : f.id = newid$)}{
        //~required event has happened\;
         \AddEventToToken($t$,$f$)\nllabel{optleastEvent}\;
         \EvaluateToken($t$)\nllabel{optevaltokenafterleast}\;
        }
}
\BlankLine
\Indm \ReceiveCut(Event $event$,Vector $cutV$,Vector $stateV$)\;\nllabel{recvCutStart}%$senderId$,
\Indp \ForEach(//~check if $t$ can be updated){stalled token $t$ at $S_i$}{
   \CopyCutIfNeeded($t, event, cutV, stateV, i$)\; 
}\nllabel{recvCutEnd}
\remove{
\BlankLine
\Indm \CopyCutDetails($t$,$cutV$)\;
\Indp
      $t.eval \leftarrow true$\; 
      $t.gcut \leftarrow cutV$\;  
      $t.depend \leftarrow cutV$\;  
      $t.target \leftarrow t.gcut[i] +1$\;
}
\caption{Optimized Routines at $S_i$}
\label{alg:optimized}
\end{algorithm}
%\remove{
\begin{algorithm}[!ht]
\DontPrintSemicolon
\SetKwFunction{ReceiveEvent}{ReceiveEvent}
\SetKwFunction{ReceiveToken}{ReceiveToken}
\SetKwFunction{EvaluateToken}{EvaluateToken}
\SetKwFunction{UpdateStalledTokens}{UpdateStalledTokens}
\SetKwFunction{SendIfNeeded}{SendIfNeeded}
\SetKwFunction{CopyCutIfNeeded}{CopyCutIfNeeded}
\SetCommentSty{small}
\SetProcNameSty{textbf}
\UpdateStalledTokens(Token $t'$)\;\nllabel{updatestallstart}
\Indp \ForEach{stalled token $t$ at $S_i$}{
   \CopyCutIfNeeded($t, t'.event, t'.cutV, t'.stateV, t.pid$)\; 
   }
\If{($t'.event.type = MSGSEND$)}{
\Indm \Indm \tcc*[l]{send cut details to message recipient process; $k$: id of the message recipient process} 
\Indp $S_{k}$.\ReceiveCut($t'.event, t'.gcut, t'.gstate$)\;
}\nllabel{updatestallend}
\BlankLine
\Indm \CopyCutIfNeeded(Token $t$, Event $event$, Vector $cutV$, Vector $stateV$, int $ignore$)\;
\Indp 
 \If(//~$t$ was waiting for update from $event$){$t.target = event$}{\nllabel{checkone}
   \ForEach(~~//copy relevant details){$j$ in $1$ to $n$ s.t. $j \neq ignore$}{
      $t.gcut[j] \leftarrow cutV[j]$\;  
      $t.depend[j] \leftarrow cutV[j]$\; 
      $t.gstate[j] \leftarrow stateV[j]$\; 
   }
   \tcc*[l]{clear stalled state; ensure no duplicate output}
         $t.stalled \la false$\;
         $t.eval \la predfalse$\;
      \If{($t.event \in cutV$)}{\nllabel{checktwo}
   \tcc*[l]{cuts are same, move on to next event}
       $t.target \leftarrow t.gcut[i] +1$\; \nllabel{nextevent}
       \If{($t.pid \neq i$)}{
         {\bf send} $t$ to $S_{t.pid}$ 
       }
      }
      \Else{                      %{$(t.target || event) \land t.event \in cut$}
   \tcc*[l]{cuts not same, resume token's computation}
        \ProcessToken($t$)\; 
      }
}
\caption{Helper Routines at $S_i$}
\label{alg:helpers}
\end{algorithm}

We do not reproduce the functions {\tt ReceiveEvent}, {\tt EvaluateToken}, {\tt ReceiveStopSignal} and {\tt SendIfNeeded} in the modified algorithm (in Algorithm~\ref{alg:optimized}), as 
they remain identical to their earlier versions. 
In the optimized algorithm an additional variable, $currentE$ - at each {\em slicer} process $S_i$, is used as a local pointer to keep track of the event $e$ for which $S_i$
is currently computing $J_B(e)$. Tokens also stores this information (with $token.event$), however  
the token $T_i$ is not always present on $S_i$. By keeping $currentE$ updated, even in absence of token $T_i$, the {\em slicer} process $S_i$ can 
delay the progress of other tokens whenever it suspects that these tokens may undergo the same $J_B(e)$ computation that is being considered by $T_i$. 
For stopping possibly duplicate computations, a flag, called $stalled$, is maintained in each token. By setting the $stalled$ flag on any token, a slicer removes the token 
from the set of $waiting$ tokens; and  
no updates are performed on tokens that are in the $stalled$ state. These modifications allow {\em slicer} processes 
to delay the computation progress on $stalled$ tokens, 
and ensure that no two tokens finish and output any two join-irreducible consistent cuts that are equal. The optimized 
algorithm also makes use of the type information of events, for identifying if an event is a send of a message (denoted 
by type $MSGSEND$) or a receive of a message (denoted by type $MSGRECV$). The modifications 
are briefly explained below:\\
\\
{\bf Algorithm~\ref{alg:optimized}}: Optimized Routines
\hrule 
\vspace{1pt}
{\bf Lines}~\ref{recvWaitStart}-\ref{recvWaitEnd} and~\ref{recvCutStart}-\ref{recvCutEnd}: If the type of an event $e \in E_i$ indicates that the event is a message receipt, 
then the algorithm stalls the token $T_i$, and sets its `target' as the message send event, $f$. This step is to incorporate 
Lemma~\ref{lem:happBefJB} in speculative manner. Whenever the corresponding {\em slicer} process of  message send event
($f$) finishes computing the $J_B(f)$, it informs $S_i$ about the computed cut. $S_i$ on receiving this cut, calls 
the helper sub-routine {\tt CopyCutIfNeeded} (shown in Algorithm~\ref{alg:helpers}) that checks 
if $e$ belongs to $J_B(f)$ and thus $J_B(e)$ computation is not needed; otherwise $S_i$ restarts the computation for $J_B(e)$.\\
{\bf Lines}~\ref{concStallStart}-\ref{concStallEnd}: These steps incorporate Lemma~\ref{lem:bothinJB} in speculative 
manner. The first condition  $t.event \not\ra currentE$ ensures that if the current computation on token $T_i$ is causally
dependent on the computation on token $t$, then $t$ is not stalled. The second condition is evaluated only if $t.event$ 
and $currentE$ are not causally related, i.e. they are concurrent. If this is the case, then the check ($newid \geq currentE \land i > t.pid$) ensures that the computation of $J_B(t.event)$  does not progress beyond the current ongoing computation on $S_i$, and performs 
symmetry breaking (so that there is no deadlock) in favor of the token/process with larger process id. This guarantees
that whenever the cuts of two concurrent events are same, only one of the tokens (with the smaller process id) 
finishes computing the cut, and thus duplicate computations are not performed. \\
{\bf Line}~\ref{updateStalledCall}: Whenever $S_i$ finishes computing the $J_B(e)$ for the event $e \in E_i$, it 
tries to update all the tokens 
that were speculatively stalled to avoid computing the same cut. The steps involved are explained next.\\
\hfill\\
{\bf Algorithm~\ref{alg:helpers}}: Helper Routines 
\hrule 
\vspace{1pt}
{\bf Lines}~\ref{updatestallstart}-\ref{updatestallend}: For each token that is stalled, either locally or at some other
{\em slicer} process, due to the event $t'.event$, update (tokens present locally) or notify (at some other process) it. The
notification to other {\em slicer} processes is performed by sending the details of the cut to them. If the stalled tokens infer, using the checks 
of lines~\ref{checkone} and \ref{checktwo}, that their cuts (if computed) and $t'.gcut$ would be same by the application of 
Lemmas~\ref{lem:happBefJB} and \ref{lem:bothinJB}, then they copy the cut details of $t'$ and move on to the next events on
their respective processes. 

\subsection{Example of Optimized Algorithm Execution}

We revisit the example presented in section \ref{sec:run_ex} for the distributed algorithm run, in order to 
show the difference in execution for the optimized 
algorithm. 
When $f$ is reported to $S_2$, the earlier version of the 
algorithm has to update the token $T_2$, and send 
it to $S_1$ in order to make the cut on $T_2$ consistent. 
The optimized algorithm uses the checks on lines~\ref{recvWaitStart}-\ref{recvWaitEnd} in
Algorithm~\ref{alg:optimized}, and determines 
that $f$ being a message-receive event, the $J_B(f)$ 
computation should not be started until the corresponding
message-send event's computation is reported to $S_2$. Thus $T_2$ would be kept in $stalled$ state, until $T_1$ 
finishes the $J_B(b)$ computation. When $J_B(b)$ computation on $T_1$ is finished, with $J_B(b) = [b,f]$, then 
as per line~\ref{updateStalledCall} of Algorithm~\ref{alg:optimized}, $S_1$ would send the information of $b$'s join-irreducible
cut to $S_2$. On receiving the cut details, $S_2$ would try to update its stalled token $T_2$, and as per lines~\ref{checkone}-\ref{nextevent}, it would infer that $J_B(f) = J_B(b)$. Thus, it would just copy the details of the cut as the 
result for $J_B(f)$, 
and move on to computing $J_B(g)$. 
\subsection{Proof of Correctness}

To prove the correctness of the optimized version of the algorithm, it suffices to show that this version 
does not introduce deadlocks, and as desired -- does not enumerate duplicate join-irreducible consistent cuts. 
\begin{thm}
\label{thm:optnodeadlock}
Algorithm~\ref{alg:optimized} cannot lead to deadlocks. 
\end{thm}
\begin{\pf}
Let us assume that the algorithm leads to a deadlock. Thus, even in presence of events required to process, a set of tokens is not able to make progress, with every token in this set being in the stalled state. There are 
two possible scenarios for this:\\ 
{\bf (a)}: The tokens are stalled such that for each token $T_i$, the event $T_i.event$, for which it is computing the 
join-irreducible cut, is concurrent to every other
token $T_j$'s event $T_j.event$. As all the tokens have unique positive ids, there is a unique minimum id. By the required
check performed at 
line~\ref{concStallStart} in Algorithm~\ref{alg:optimized}, such a token can never be stalled. Thus, we have a contradiction. \\
{\bf (b)}: All the tokens are stalled such that for at least one pair of tokens $T_i$ and $T_j$, $T_i.event \ra T_j.event$. 
In such a case, by the check performed at 
line~\ref{recvWaitStart} in Algorithm~\ref{alg:optimized}, $T_i$'s token can never be stalled at $S_j$. Thus $T_i$ must be stalled 
at some other process $S_k$ such that $T_i.event$ is concurrent with $T_k.event$. But either $i < k$ or $i > k$, and in 
either case, by applying the result of case (a) above, we are guaranteed that $T_i$ would eventually make progress. Thus
$T_i$ would not remain stalled forever. Again, we have a contradiction.  
\end{\pf}
\begin{thm}
\label{thm:nodupsfinish}
If $J_B(e) = J_B(f)$ for two distinct events $e$ and $f$, then only one of these cuts is enumerated in  Algorithm~\ref{alg:optimized}. 
\end{thm}
\begin{\pf}
Given that $e \neq f$, then 
either: (a) $e || f$ (both are concurrent), or $e \ra f$ (assume without loss of generality). If $e || f$, then 
as per line~\ref{concStallStart} one of their tokens would be stalled on the other process. Again, without loss of generality assume
$f$'s token was stalled on $e$'s process. Thus, given that there can be no deadlocks, the $J_B(e)$ computation 
would eventually finish, and update $f$'s token to indicate that it should move to the successor of $f$ (as per the steps
in Algorithm~\ref{alg:helpers}). Hence, $J_B(f)$ would not be enumerated.\\
If $e \ra f$, then $f$'s token would be stalled on its own process, until the completion of $J_B(e)$ 
computation, upon which $f$'s token would be made to move to the event that is the successor of $f$. Again $J_B(f)$ would not be enumerated.   
\end{\pf}

\remove{
\subsection{Analysis}

The upper bounds on space and computation complexities of the optimized distributed  algorithm remain the same for 
the worst case scenarios.
 The optimized version also performs $O(n|E|)$ work/process when averaged across all the processes. The upper bound for space required on any 
process is $O(|E_i||S|)$, where $|S|$ denotes the number of bits required 
to capture information of a local state.  
}

\subsection{Analysis}
\label{sec:analysis}
Each token $T_i$ processes every event $e \in E_i$ once for computing its $J_B(e)$.
If there are $|E|$ events in the system,
then in the worst case $T_i$ does $O(n|E|)$ work, 
because it takes $O(n)$ to process
one event. We are assuming here that evaluation of $B$ takes $O(n)$ time
given a global state. There are $n$ tokens in the system,
hence 
 the total work performed is $O(n^2|E|)$.
Since there are $n$ slicing processes and $n$ tokens, the average work 
performed is $O(n|E|)$ per process. In comparison, 
the centralized algorithm (either online or offline) 
requires the {\em slicer} process to perform $O(n^2|E|)$
work.

%Table \ref{tbl:bounds} presents the complexity bounds on total work, work/slicer, message, message-bits, and consumed heap-space 
 Let $|S|$ be the maximum number of bits required to represent a local 
state of 
a process.     
The actual value of $|S|$ is subject to the predicate under consideration, as the resulting number/type of the variables to capture 
the necessary information for predicate detection depends on the predicate. 
The centralized online algorithm requires $O(|E||S|)$ 
space in the worst case; however it is important to 
notice that all of this space is required on a single 
(central slicer) process. For a large computation, 
this space requirement can be limiting. The distributed algorithm proposed above only consumes $O(|E_i||S|)$ space per slicer.
Thus, we have a reduction of $O(n)$ in per slicer space consumption. 

The token can move at most once per event. Hence, in the worst case
the message complexity is $O(|E|)$ per token. Therefore, the message
complexity of the distributed algorithm presented here is $O(n|E|)$ total for all tokens.
The message complexity of the centralized 
online slicing algorithm is $O(|E|)$ because all the 
event details are sent to one (central) slicing process. 
However, for conjunctive predicates, it can be observed that the message 
complexity of the optimized version of the distributed algorithm is also  $O(|E|)$. With speculative stalling 
of tokens, only unique join-irreducible cuts are computed. This means 
that for conjunctive predicates, a token only leaves (and returns to) 
$S_i$, $O(|E_i|)$ times. As there are $n$ tokens, the overall message
complexity of the optimized version for conjunctive predicates is $O(|E|)$. 

\remove{
\begin{table*}[!htb]
\caption{Comparison of Complexity Bounds of Online Algorithms}
\centering
    \begin{tabular}{ | c | c | c | c | c | c | }
        \hline
            Algorithm & Total Work & Work/Slicer & Messages & Total Bits & Space/Slicer \\ \hline
            Centralized & {\em O}($n^2$$|E|$) & {\em O}($n^2$$|E|$) & {\em O}($|E|$) & {\em O}($|E||S|$) & {\em O}($|E||S|$)\\ \hline
            Distributed (this paper) & {\em O}($n^2$$|E|$)& {\em O}($n|E|$) & {\em O}($n|E|$) & {\em O}($n|E||S|$) & {\em O}($|E_i||S|$) \\ \hline
\end{tabular}
\label{tbl:bounds}
\end{table*}
}
\remove{
\begin{table}[!h]
\caption{Comparison of Work for Online Algorithms}
\centering
    \begin{tabular}{ | l | l | l |}
        \hline
            Algorithm &  \\ \hline
            Centralized & \\ \hline
            Distributed \\(this paper) & \\ \hline
\end{tabular}
\label{tbl:workBounds}
\end{table}
}

%\section{Example}
%\input{example.tex}
%\section{Proof of Correctness}
\section{Evaluation \& Discussion}
As indicated by the analysis of the distributed algorithm in Section \ref{sec:analysis}, the distributed algorithm 
reduces the worst case complexity for the total work per slicer process by an order of magnitude. 
\remove{However, the worst case message 
and message-bit complexities  for the distributed algorithm appear to be much higher than those of the centralized algorithm.
This seems a bit counter intuitive as one primary motivation for a distributed algorithm is to reduce the maximum message loads on 
a single process.
}To better understand the actual gains of the distributed algorithm, we implement both centralized 
and distributed algorithms, and evaluate their performance based on the experimental results for slicing on the same set of computations.  

The experiments were performed on a 64 bit, 8 processor (2.3 GHz) machine with 4GB memory, running Linux 3.2.0-32 kernel. Each process
is allowed to run till it executes its local program counter to a fixed upper-bound. For the reported results the local program counter upper-bound was set to $100$. Message activity 
was decided in a randomized manner. After each local state change the processes could 
send a message, with a probability of $0.8$, to a randomly chosen process.  The number of processes in the computation was varied from 2 to 10. We 
monitor the total object sizes of the centralized slicer, and each of the distributed slicer processes while they find 
the slice for an ongoing computation. We also monitor 
the total number of messages received by all the slicer processes during the online computations. For any distributed slicer $S_i$,
the reported number of messages includes the token messages received by it from 
other slicers. We present a comparison of the centralized and distributed algorithms in terms of maximum values of both the consumed space and total messages per slicer process. 
\remove{
Evaluation of the computational overhead of running the distributed online
slicing algorithm, in comparison to the actual processing time 
taken by the computation itself. In addition, the experimental evaluation 
of the overheads of the centralized and the distributed online algorithms. 
}

Fig. \ref{fig:conj}{\subref{fig:conjMemResults} plots the ratio of maximum space consumed by any distributed slicer object 
and the maximum space used by the centralized slicer (for the same computation), against the number of processes in the computation. The space consumption is evaluated at pre-determined check points that are symmetric for both 
centralized and distributed algorithms. 
Fig. \ref{fig:conj}{\subref{fig:conjMsgResults} presents the maximum number of messages received by centralized slicer and that 
received by any distributed slicer for the same instances of computations.
\remove{
We also monitor the  CPU consumption by the slicer processes, as an additional overhead on top of the ongoing computations. 
The CPU overhead values for the centralized algorithm, and the max overhead by any distributed slicer for all our evaluations are 
relatively very small, and very similar. Due to space limitations, we skip reporting these details here.}
\begin{figure*}
\centering
\subfloat[{\label{fig:conjMemResults}}Comparison of Maximum Space per Slicer]
{\includegraphics[angle=-90, scale=0.4]{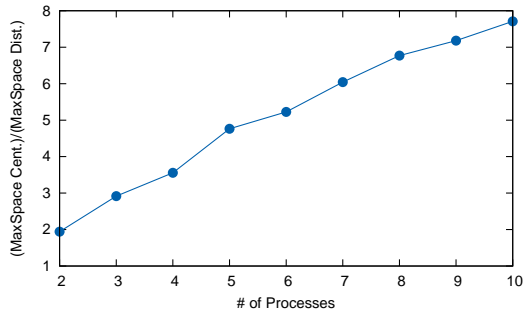}}
\qquad
\subfloat[{\label{fig:conjMsgResults}}Max. messages per Slicer]
{\includegraphics[angle=-90, scale=0.4]{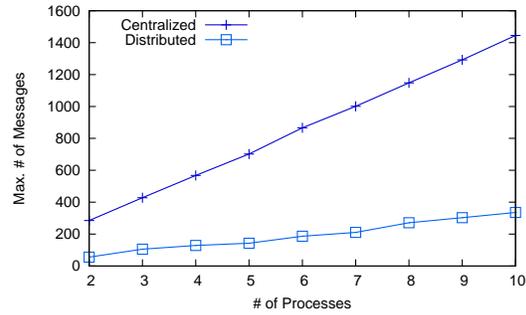}}
\caption{Comparison of Memory Usage and Total Messages for Conjunctive Predicate}
\label{fig:conj}
\end{figure*}
 
\section{Related Work}

The distributed algorithm presented in this paper constructs the slice of a distributed computation with respect to a regular state based predicate. The constructed slice can then be used to determine if some consistent cut of the computation satisfies the predicate. This is referred to as the problem of detecting a predicate under \emph{possibly} modality~\cite{CooMar:1991:WPDD}.
In~\cite{CooMar:1991:WPDD}, a predicate is detected by exploring the complete lattice of consistent cuts in a breadth first manner. Alagar et. al.~\cite{AlaVen:2001:TSE} use a depth first traversal of the computation lattice to reduce space complexity. The algorithms in~\cite{CooMar:1991:WPDD} and~\cite{AlaVen:2001:TSE} can handle arbitrary predicates, but in general have exponential time complexity. In contrast, the slicing algorithm presented in this paper for a {\em regular} predicate has polynomial time complexity. 

In this paper we assume a static distributed system. Predicate detection algorithms have been proposed for dynamic  systems~(e.g.~\cite{WanMay+:2010:ICPADS,DarMay+:2006:PDS,DhaSri+:1997:PC,JohMit:2009:ICDCS,WanMay+:2009:ICPADS}), where processes may leave or join. However, these algorithms detect restricted classes of predicates like stable predicates and conjunctive local predicates, which are less general than regular predicates.
In computation slicing, we analyze a single \emph{trace} (or execution) of a distributed program for any violation of the program's specification. Model checking~(cf.~\cite{clarke00}) is a formal verification technique that involves determining if (all traces of) a  program meets its specification. Model checking algorithms conduct reachability analysis on the state space graph, and have a time complexity that is exponential in number of processes. 

Partial order methods~(cf.~\cite{God:1996:SV}) aim to alleviate the state-explosion problem by minimizing the state space for predicate detection. This is done by exploring only a subset of the interleavings of concurrent events in a computation, instead of all possible interleavings. However, predicate detection algorithms based on partial order methods still have exponential time complexity, in the worst case. In this paper, the focus is on generating the slice with respect to a predicate. Partial order methods such as~\cite{StoUnn+:2000:CAV} can be used in conjunction with slicing to explore the state space of a slice in a more efficient manner~\cite{MitGar:2003:ICDCS}. 

The work presented in this paper is related to runtime verification~(cf.~\cite{Leucker2009293}), which involves analyzing a run of a program to detect violations of a given correctness property. 
The input program is instrumented and the trace resulting from its execution is examined by a monitor that verifies its correctness.
Some examples of runtime verification tools are Temporal Rover~\cite{drusinsky00}, Java-MaC~\cite{kim01}, JPaX~\cite{havelund01}, JMPaX~\cite{sen05}, and jPredictor~\cite{chen-serbanuta-rosu-2008-icse}. 
The Temporal Rover, Java-Mac and JPaX tools model the execution trace as a total order of of events, which is then examined for violations.
In the JMPaX and jPredictor tools, as in our algorithm, the trace is modeled as a partial order of events. A lattice of consistent cuts of the computation is then generated, which is searched by the monitor. Further, these tools generate states not observed in the current trace, to predict errors that may occur in other runs, thereby increasing the size of the computation lattice. 
Chen et al.~\cite{CheRos:2007:CAV} note that computation slicing can be used to make tools like jPredictor more efficient by removing redundant states from the lattice.
All of these tools are centralized in nature, where the events are collected at a central monitoring process. Sen et al.~\cite{SenVar+:2004:ICSE} present a decentralized algorithm that monitors a program's execution, but can only detect a subset of safety properties. The distributed algorithm presented by Bauer et al.~\cite{BauFal:2012:FM} can handle a wider class of predicates, but requires the underlying system to be synchronous.

%%Lines that can be included.
%Detection of regular predicates under \emph{definitely} modality was shown to be coNP-Complete in~\cite{DBLP:conf/atva/Huang09}.
%Huang~\cite{Huang:2011:ICIS} presents two centralized algorithms for detecting regular predicates under definitely modality, for a computation in which every global state is consistent. We do not make any such assumptions about the underlying computation. 
%Predicate detection in systems with faulty processes has also been studied~(cf.~\cite{GarPle:2005:DCS}). However, the class of predicates that can be detected by such algorithms is also very restricted~\cite{GarPle:2001:S-SS}.
%Sen et al.~\cite{SenVar+:2004:ICSE} present a decentralized algorithm that monitors a program's execution, but can only detect a subset of safety properties.

\label{sec:conc}
\section{Conclusion}
In this paper, we presented a distributed online algorithm for performing {\em computation slicing}, a technique
to abstract the computation with respect to a regular 
predicate. The resulting abstraction ({\em slice}) is usually much smaller, sometimes exponentially, in size. For regular
predicates, by detecting the predicate only on the abstracted computation, one is guaranteed to detect the predicate in the 
full computation, which leads to an efficient detection mechanism.  
\remove{
The family of regular predicates contains 
many predicates that are of interest for the runtime verification domain, especially for detecting constraint violations. 
}
By distributing the task of abstraction among all the 
processes, our distributed algorithm reduces the space required, as well as computational load on a single process by a factor of $O(n)$.
We also presented an optimized version of the distributed algorithm that does not perform redundant computations, and requires 
reduced number of messages.
The results of experimental evaluation (available in extended version of this paper at \cite{arxfull}) compare the performance of our distributed 
algorithm with that of the existing centralized algorithm.

\bibliographystyle{IEEEtran}
\bibliography{citations_a,slicing_a,bibliography/citations,bibliography/slicing}
\newpage
\remove{
\appendix
\input{latticefig.tex}
}

\end{document}